%% file: mnras_template.tex
\documentclass[fleqn,usenatbib]{mnras}

\usepackage{newtxtext,newtxmath}

\usepackage[T1]{fontenc}
\usepackage{tikz} 
\usetikzlibrary{shapes.geometric, arrows}
\tikzstyle{arrow} = [thick,->,>=stealth]
\usepackage{dirtytalk}
\DeclareRobustCommand{\VAN}[3]{#2}
\let\VANthebibliography\thebibliography
\def\thebibliography{\DeclareRobustCommand{\VAN}[3]{##3}\VANthebibliography}

\usepackage{graphicx}	
\usepackage{amsmath}	
\usepackage{lscape}
\usepackage{booktabs}
\usepackage{longtable}
\usepackage{supertabular}
\usepackage{subcaption}

\usepackage{ulem}
\usepackage{xspace}
\usepackage{pifont}
\usepackage[para,online,flushleft]{threeparttable}
\newcommand{\cmark}{\ding{51}}%
\newcommand{\xmark}{\ding{55}}%

\newcommand{\vfrag}[0]{$v_\mathrm{frag}$\xspace}
\newcommand{\dustpy}{\textsc{DustPy}\xspace}

\newcommand{\bqm}{\textbf{?}}

\title[Turbulence and dust fragility in discs]{Turbulence and dust fragility in protoplanetary discs}

\author[Tong, Alexander \& Rosotti]{
Simin Tong$^{1}$ \thanks{E-mail: st547@leicester.ac.uk, astro.stong@gmail.com},
Richard Alexander$^{1}$
and Giovanni Rosotti$^{2}$
\\
$^{1}$School of Physics \& Astronomy, University of Leicester, University Road, Leicester, LE1 7RH, UK\\
$^{2}$Dipartimento di Fisica, Università degli Studi di Milano, Via Celoria, 16, 20133, Milano, Italy\\
}

\date{Accepted XXX. Received YYY; in original form ZZZ}

\pubyear{2025}

\begin{document}
\label{firstpage}
\pagerange{\pageref{firstpage}--\pageref{lastpage}}
\maketitle

\begin{abstract}
Dust growth from micron- to planet-size in protoplanetary discs involves multiple physical processes, including dust collisions, the streaming instability, and pebble accretion. Disc turbulence and dust fragility matter at almost every stage. Previous studies typically vary one of them while fixing the other, failing to provide a complete picture. Here, we use analytical models and numerical dust evolution models \dustpy to study the combinations of gas turbulence and dust fragility that can reproduce multi-wavelength ALMA observables. We find that only appropriate combinations -- fragile dust (\vfrag= 1--$2~\mathrm{m~s^{-1}}$) in discs with viscous $\alpha=10^{-4}$ or resilient dust (\vfrag= 6--$10~\mathrm{m~s^{-1}}$) in discs with viscous $\alpha=10^{-3}$ -- can reproduce observations. Our result is robust to two widely used opacities (DSHARP and Ricci opacities). Regardless of the strength of disc turbulence, reproducing observations requires observed dust rings to be optically thick at $\lambda=1.3$ and $3$ mm. As only small dust can be lifted above the midplane to reach the emitting layers, SED analysis probably yields lower limits on the maximum grain sizes. We highlight the challenge of creating detectable dust rings at large radii when incorporating bouncing in models, and the need for earlier formation of dust rings at smaller radii to reproduce the decreasing ring brightness with radius observed across ALMA wavelengths. 

\end{abstract}

\begin{keywords}
accretion, accretion discs -- protoplanetary discs -- submillimetre: planetary systems
\end{keywords}



\section{Introduction}
\input{mainbody/intro}

\section{Analytical analysis}\label{sec:analytical}
\input{mainbody/analytical}

\section{Numerical analysis}\label{sec:numerical}
\input{mainbody/numerical}
\subsection{Methods}\label{subsec:numerical_methods}
\input{mainbody/numerical_methods}
\subsection{Results}\label{subsec:numerical_results}
\input{mainbody/numerical_results}

\section{Discussion}\label{sec:discussion}
\input{mainbody/discussion}
\section{Conclusion}\label{sec:conclusion}
\input{mainbody/conclusion}

\section*{Acknowledgements} 
We acknowledge Jane Huang and Enrique Macías for providing multi-wavelength observational data for GM Aur and HD 169142. ST is thankful Yangfan Shi for interesting discussions during conducting this research. ST acknowledges Richard Grant for his assistance with resolving computer issues at the final stage of preparing the manuscript. ST acknowledges the University of Leicester for a Future 100 Studentship. RA acknowledges funding from the Science \& Technology Facilities Council (STFC) through Consolidated Grant ST/W000857/1. GR acknowledges support from Fondazione Cariplo, grant No. 2022-2017. This project has received funding from the European Research Council (ERC) under the European Union’s Horizon Europe Research \& Innovation Programme under grant agreement no. 101039651 (DiscEvol). Views and opinions expressed are however those of the author(s) only, and do not necessarily reflect those of the European Union or the European Research Council Executive Agency. Neither the European Union nor the granting authority can be held responsible for them. This work has been supported by COST Action CA22133 – The birth of solar systems (PLANETS), funded by the European Cooperation in Science and Technology (COST). This research used the ALICE High Performance Computing Facility at the University of Leicester. This paper makes use of the following ALMA data: 
ADS/JAO.ALMA\#2015.1.00490.S, ADS/JAO.ALMA\#2016.1.01158.S, ADS/JAO.ALMA\#2017.1.01151.S,
and ADS/JAO.ALMA\#2018.1.01230.S. 
ALMA is a partnership of ESO (representing its member states), NSF (USA) and NINS (Japan), together with NRC (Canada), NSTC and ASIAA (Taiwan), and KASI (Republic of Korea), in cooperation with the Republic of Chile. The Joint ALMA Observatory is operated by ESO, AUI/NRAO and NAOJ. For the purpose of open access, the author has applied a Creative Commons Attribution (CC BY) licence to the Author Accepted Manuscript version arising from this submission.

\section*{Data Availability}

Data generated in simulations and codes reproducing figures in this work are available on reasonable request to the corresponding author. Data for Fig. 2 is obtained from \url{https://almascience.nrao.edu/aq/} and private communications with Enrique Macías for HD 169142, and from \url{https://zenodo.org/records/3628656} for GM Aur.
This work made use of \textsc{DustPy} \citep{2022ApJ...935...35S}, \textsc{DustyPyLib} \citep{2023ascl.soft10005S}, \textsc{RADMC-3D} \citep{2012ascl.soft02015D}, \textsc{Jupyter} \citep{jupyter}, \textsc{Matplotlib} \citep{matplotlib}, \textsc{Numpy} \citep{numpy, numpy2}, \textsc{Scipy} \citep{scipy2020}, \textsc{Astropy} \citep{astropy2018}, \textsc{Pandas} \citep{pandas2010, pandas2020}, \textsc{CASA} \citep{2022PASP..134k4501C} and \textsc{GoFish} \citep{Teague2019}.



\bibliographystyle{mnras}
\bibliography{reference} 




\appendix
\input{appendix/appendix1}

\bsp	

\end{document}

%% file: mainbody/intro.tex
Dust grains in protoplanetary discs are the building blocks of planets. The growth of interstellar dust to millimetre/centimetre-pebbles, then to kilometre-planetesimals, and ultimately to planets has been extensively studied in protoplanetary discs \citep[e.g.][and references therein]{2023ASPC..534..717D}. 

Theoretically, dust growth over several orders of magnitude involves multiple physical mechanisms. Initially, micron-sized dust collides and sticks together to form millimetre/centimetre-sized aggregates \citep[][references therein]{2024ARA&A..62..157B}. These pebbles concentrate to enhance the dust-to-gas ratio and trigger the streaming instability \citep{2005ApJ...620..459Y}. Further growth requires pebble accretion \citep{2017AREPS..45..359J} and planetesimal accretion \citep{2024arXiv241211064A}. 

The disc turbulence level and dust fragility are relevant to every stage of these processes. Turbulence plays a key role in regulating the efficiency of gas and angular momentum transport, in addition to other mechanisms such as magneto-hydrodynamic winds \citep{1982MNRAS.199..883B}. 
High turbulence tends to shorten the disc lifetime by enhancing the inward transport of both gas and small dust that is well coupled to the gas \citep{1986Icar...67..375N}. Turbulence also determines dust growth \citep{2009A&A...503L...5B}, dust radial diffusion \citep{2007Icar..192..588Y} and vertical distributions \citep{1995Icar..114..237D}. The interplay among dust size distributions, spatial distributions and turbulence can further affect the onset of the streaming instability \citep[e.g.][]{2021ApJ...919..107L, 2024ApJ...969..130L}. Notably, subsequent processes such as pebble accretion are also found to be more efficient in a low turbulence environment \citep{2025MNRAS.540..165N}. The dependence highlights the importance of a better understanding of disc turbulence levels. 

Over the last decade, various approaches have been developed to constrain turbulence levels. \citet{2004ApJ...603..213C, 2015ApJ...813...99F} measure disc turbulence through non-thermal contributions to the line broadening, while \citet{2018ApJ...869L..46D, 2025arXiv250313818Y} constrain disc turbulence from dust radial diffusion around dust rings. \citet{2022ApJ...930...11V, 2025A&A...697A..64V} estimate disc turbulence from dust vertical settling. A broader review of turbulence measurement is given by \citet{2023NewAR..9601674R}. Though it can vary from disc to disc \citep{2020ApJ...895..109F}, recent studies infer values of $\alpha \lesssim 10^{-3}$ in the outer regions of these discs ($r\gtrsim 10$ au). 

Dust fragility is the property which characterises the likelihood that dust coagulates rather than fragments when particles collide, and is usually parametrized in terms of the threshold velocity for fragmentation. It has been investigated through laboratory experiments, observations and numerical simulations, though results remain inconclusive. Some experiments indicate uniformly fragile dust ($\sim 1~\mathrm{m~s^{-1}}$) across the disc \citep[e.g.][]{2019ApJ...873...58M}, and fragile dust is also supported by non-polarimetric and polarimetric observations \citep[e.g.][]{2024NatAs...8.1148U} and by the estimated pebble flux \citep[e.g.][]{2025MNRAS.537..831W}. On the other hand, some experiments \citep{2005Icar..178..253W}, observations \citep{2025MNRAS.538.2358S} and numerical simulations \citep[e.g.][]{2024A&A...688A..81D} prefer resilient dust ($\sim 10~\mathrm{m~s^{-1}}$).

A typical approach to estimate turbulence from observations is fitting steady-state analytical models to observed structures \citep[e.g.][]{2018ApJ...869L..46D, 2020MNRAS.495..173R}. However, these models often yield best-fit parameters that are sensitive to assumptions and may fail to reproduce other observables of the given disc. For example, a disc with turbulence of $\alpha\sim10^{-3}$ in the outer disc may struggle to form millimetre dust when dust is fragile \citep{2025MNRAS.537.3525T}, despite observational inference of millimetre dust in several discs \citep[e.g.][]{2019ApJ...883...71C, 2021ApJS..257...14S}.

In this paper, we aim to constrain combinations of disc turbulence and dust fragility that can simultaneously reproduce multiple observables from recent multi-wavelength observations with disc \textit{evolution} models, in addition to estimating them from an analytical solution. We use \dustpy \citep{2022ApJ...935...35S}, a Python package that incorporates gas and dust transport, as well as the dust coagulation algorithm, for the numerical studies. The paper is organized as follows: we present the analytical analysis in Section \ref{sec:analytical}, numerical analysis in \ref{sec:numerical}. We discuss results in Section \ref{sec:discussion} and summarize our conclusions in Section \ref{sec:conclusion}.

%% file: mainbody/analytical.tex
In the micron to millimetre size range, dust growth can be hindered by several mechanisms, including electrostatic repulsion \citep{2009ApJ...698.1122O}, bouncing \citep{2010A&A...513A..57Z}, radial drift \citep{1977MNRAS.180...57W, 2008A&A...480..859B} and fragmentation \citep{2008ARA&A..46...21B}. Assuming neutral dust and neglecting bouncing, which will be discussed in Section \ref{subsec:discussion_dust_prop}, it leaves only the drift and fragmentation barriers. Radial drift limits dust growth when particles are decoupled from the gas and drift inwards before they grow larger, while fragmentation typically limits dust growth for fragile dust and in regions where dust becomes concentrated, such as in the inner discs and effective dust traps, which can effectively trap dust locally, yielding more frequent collisions.

\subsection{Methods}\label{subsec:ana_method}
Pressure bumps are effective dust traps \citep{1977MNRAS.180...57W}. The locally high dust concentration enhances the frequency of dust collisions. This makes dust fragmentation the dominant limiting mechanism for dust growth around pressure bumps. Under the assumption that the maximum dust size is determined by fragmentation around pressure bumps, we can estimate disc turbulence and dust fragility from other already known disc properties.

The fragmentation-limited maximum dust size can be obtained by equating the relative collision velocity to the fragmentation velocity \vfrag, which is a threshold between constructive and destructive collisions \citep{2012A&A...539A.148B}.
\begin{equation}\label{eq:a_max_frag}
    a_\mathrm{max, frag}= \frac{2}{3\pi} \frac{\Sigma_g}{\rho_s \alpha_\mathrm{t}}\bigg(\frac{v_\mathrm{frag}}{c_s}\bigg)^2,
\end{equation}
or 
\begin{equation}\label{eq:st_max_frag}
    \mathrm{St}_\mathrm{frag}=\frac{1}{3\alpha_t}\bigg(\frac{v_\mathrm{frag}}{c_s}\bigg)^2
\end{equation}
in Stokes number, which is defined as the ratio of the dust stopping time-scale to the orbital time-scale (see Equation \ref{eq:stokes}). When deriving Equations \ref{eq:a_max_frag} and \ref{eq:st_max_frag}, the assumption that turbulent motion dominates the relative velocity $\Delta v=\sqrt{3\alpha \mathrm{St}}~c_s$ \citep{2007A&A...466..413O} is taken. This is true until turbulence becomes very low ($\alpha \sim 10^{-5}$).

This equation has been previously used to estimate the turbulence level in \citet{2020MNRAS.495..173R, 2023ApJ...959L..15Z, 2024A&A...682A..32J} and \citet{2025A&A...697A..64V}, where the fragmentation velocity \vfrag is mostly fixed to a few values. Various approaches are taken to other relevant parameters. \citet{2024A&A...682A..32J} adopt modified gas surface densities from multi-wavelength observations \citep{2020MNRAS.493L.108B, 2021ApJS..257....5Z}. \citet{2020MNRAS.495..173R, 2025A&A...697A..64V} use $\alpha/\mathrm{St}$ constrained independently from other methods. 

In this work, we estimate the turbulence level and dust fragility together over $\alpha=10^{-4}$--$10^{-3}$, which is slightly lower than the upper limits estimated from line broadening \citep[a few $10^{-3}$, e.g.][]{2015ApJ...813...99F, 2018ApJ...864..133T} and treat the fragmentation velocity, which is a sensitive parameter to disc turbulence, as a free parameter. We substitute parameters derived from high-quality MAPS data into Equation \ref{eq:a_max_frag}. These include 
\begin{itemize}
    \item the maximum dust sizes $a_\mathrm{max}$ at dust rings; this is measured by \citet{2021ApJS..257...14S} using multi-wavelength SED forward-modelling when considering both absorption and scattering; we adopt $a_\mathrm{max}$ from their power-law fit, given by $a_\mathrm{max}=a_\mathrm{10}\cdot(R/10)^{-b}$, where $a_\mathrm{10}$ is the maximum grain size at 10 au, $b$ is the fitted slope, $R$ is the radius in units of au. The values of $a_\mathrm{10}$ and $b$ are provided in their Table 2.
    \item the gas surface densities; this is derived from rotation curve fitting \citep{2024A&A...686A...9M}, which assumes a self-similar gas surface density (Equation \ref{eq:sigma_gas}) and fits the stellar mass $M_*$, disc mass $M_d$ and disc characteristic radius $R_c$.
    \item the sound speed; this is estimated with reported stellar parameters in MAPS \citep{2021ApJS..257....1O}, following Equation 5 in \cite{2018ApJ...869L..46D}.
\end{itemize}

The only additional parameter that is not based on MAPS data is the dust bulk density $\rho_s$, which is taken to be $1.67~\mathrm{g~cm^{-3}}$ from the \textsc{DSHARP} opacities \citep{2018ApJ...869L..45B}. This is also the opacity used to derive the maximum dust sizes in \citet{2021ApJS..257...14S}. If we consider moderately porous dust with porosity $f=0.7$-$1$ \citep{2019ApJ...885...52T}, this allows a range of dust bulk densities, varying from $1.17$ to $1.67~\mathrm{g~cm^{-3}}$. Therefore, the assumed dust bulk density reflects quite a wide range of possible combinations of the dust bulk density and porosity. We summarize parameters discussed above in Table \ref{tb:stellar_params}.

We apply this method to three out of five MAPS targets that exhibit relatively prominent dust rings (IM Lup excluded) and that have good rotation curve fitting (AS 209 excluded). For each target, we focus on its two most prominent dust rings. The relation between turbulence levels and dust fragility is shown in Fig. \ref{fig:ana_sol_vfrag}.

\subsection{Results}

The dust fragility is positively correlated with the disc turbulence level \citep[e.g.][]{2022A&A...661A..66Z}. Five out of six dust rings across three discs consistently suggest fragile dust with \vfrag $\lesssim 4~\mathrm{m~s^{-1}}$ when turbulence is low ($\alpha \sim 10^{-4}$). More turbulent discs require more resilient dust to produce millimetre-sized dust, since high turbulence boosts the relative collision velocity and can destroy fragile dust more easily before it grows to millimetre sizes. On the contrary, resilient dust in low turbulence discs grows rapidly through collisions, with most dust mass concentrated into larger particles. These large particles typically have lower opacities, making it difficult to reproduce the observed millimetre fluxes. This indicates only appropriate combinations of \vfrag and $\alpha$ can produce dust sizes inferred from observations.

The second ring R166 in MWC 480 requires higher fragmentation velocities, i.e. more resilient dust, than the other five dust rings over all ranges of turbulence levels explored here. We attribute this to its location well beyond the disc characteristic radius, which is 128 au\footnote{The gas surface density drops significantly beyond the characteristic radius. This requires a larger \vfrag to grow grains to the same $a_\mathrm{max,frag}$ (Equation \ref{eq:a_max_frag})}. The trend that dust rings located further out tend to require higher \vfrag is generally applied for any dust rings in three discs, especially when $R\gtrsim R_c$. The increasing \vfrag along the radius is discussed in Section \ref{subsec:discussion_increasing_vfrag}.

\begin{figure}
    \centering
    \includegraphics[width=1.\linewidth]{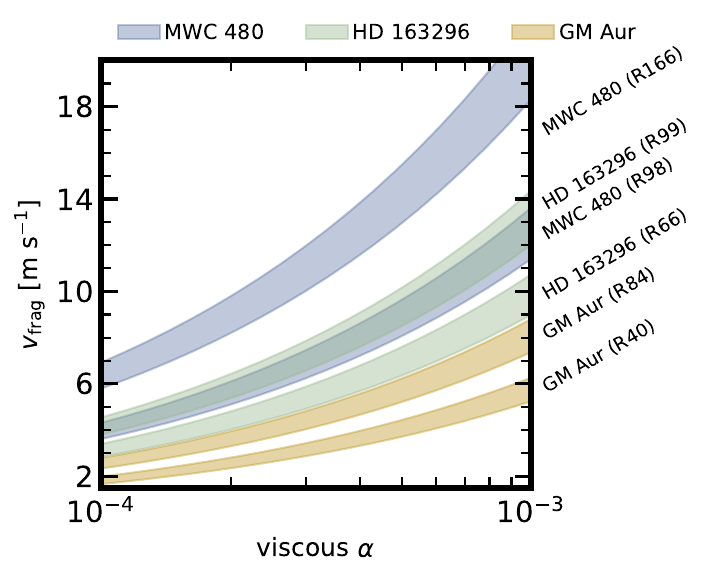}
    \caption{Correlations of disc turbulence and dust fragility for two most prominent dust rings of three MAPS samples (GM Aur, HD 163296 and MWC 480). Each ribbon represents a dust ring and is labelled with the disc name and the radius of the dust ring (R plus the radius in parentheses). The shaded regions arise from consideration of dust porosity, with is not consistently accounted for when deriving $a_\mathrm{max,frag}$ \citep{2021ApJS..257...14S} but included as demonstrated in Section \ref{subsec:ana_method}.}
    \label{fig:ana_sol_vfrag}
\end{figure}

We note that the exact \vfrag inferred from a given turbulence level in Fig. \ref{fig:ana_sol_vfrag} depends on several parameters (Equation \ref{eq:a_max_frag}) and many of which are not well constrained\footnote{The uncertainties in the stellar mass, disc mass and characteristic radius from the rotation curve fitting are relatively small, and only ranging from a few thousandths to around one per cent. We therefore do not include their contributions in Figure \ref{fig:ana_sol_vfrag}.}, especially the maximum dust size around the dust ring $a_\mathrm{max,frag}$. The derivation of $a_\mathrm{max,frag}$ from the SED analysis depends on the opacity model and the dust size distribution. The opacity model requires knowledge of dust porosity and compositions, neither of which is well determined. Even when a porosity value is assumed here (Section \ref{subsec:ana_method}), it is not treated consistently, as porosity is not originally considered in \citet{2021ApJS..257...14S}. The dust size distribution is also simplified in practice and it is typically assumed to be a power law with a fixed slope ($n(a)\propto a^{-q}$, and $q=2.5$--$3.5$) for the entire disc. Moreover, the SED analysis also assumes optically thin discs. The implication of this assumption is discussed in Section \ref{subsec:discussion_increasing_vfrag}.

\begin{table*}
\centering
\caption{Summary of stellar and disc parameters for three MAPS discs. Columns 2-4 are for stellar masses, stellar luminosities and effective temperatures, respectively. Columns 5 and 6 are for disc masses and characteristic radii, respectively. Column 7 is for the locations of dust rings.}\label{tb:stellar_params}
\begin{threeparttable}
\begin{tabular}{cccccccc}
\hline
\hline
(1) & (2) & (3) & (4) & (5) & (6) & (7) & (8) \\
Target & $M_*^{a}$ & $L_*$ & $T_\mathrm{eff}$& $M_d$ & $R_c$ & Dust rings & Ref. \\
 & [$M_\odot$] & [$L_\odot$] & [K] & [$M_\odot$] &[$\mathrm{au}$] & [au/au] & \\
 \hline
GM Aur  & $1.128\pm0.002$ & 1.2 & 4350& $0.118\pm0.002$ & $96\pm1$ & 40/84 & 1, 2, 3, 1, 1, 4\\
HD 163296 & $1.948\pm0.002$ & 17.0 & 9332 & $0.134\pm 0.001$ &$91\pm1$ & 67/100 & 1, 5, 5, 1, 1, 6\\
MWC 480 & $2.027\pm0.002$ & 21.9 & 8250 & $0.150\pm0.002$  & $128\pm1$ & 98/166 & 1, 7, 7, 1, 1, 8 \\
\hline
\end{tabular}
\begin{tablenotes}
    a. The stellar masses are taken from \citet{2024A&A...686A...9M}, but they are nearly equivalent to those derived dynamically from \citet{2021ApJS..257...18T}.\\
    References: 1. \citet{2024A&A...686A...9M}; 2. \citet{2018ApJ...865...37M}; 3. \citet{2010ApJ...717..441E}; 4. \citet{2020ApJ...891...48H}; 5. \citet{2015MNRAS.453..976F}; 6. \citet{2018ApJ...869L..42H}; 7. \citet{2009A&A...495..901M}; 8. \citet{2021ApJS..257...14S}.
   \end{tablenotes}
\end{threeparttable}
\end{table*}

%% file: mainbody/numerical.tex
The correlations between the turbulence level and dust fragility found in the last section motivate us to test various combinations of \vfrag and $\alpha$ in dust evolution models to investigate which combinations can simultaneously reproduce \textit{multiple} observables in disc continuum observations other than the inferred maximum particle sizes $a_\mathrm{max}$. We note here that the following analysis is not aiming for any specific discs but for general Class II protoplanetary discs.

\begin{figure}
    \centering
    \includegraphics[width=0.9\linewidth]{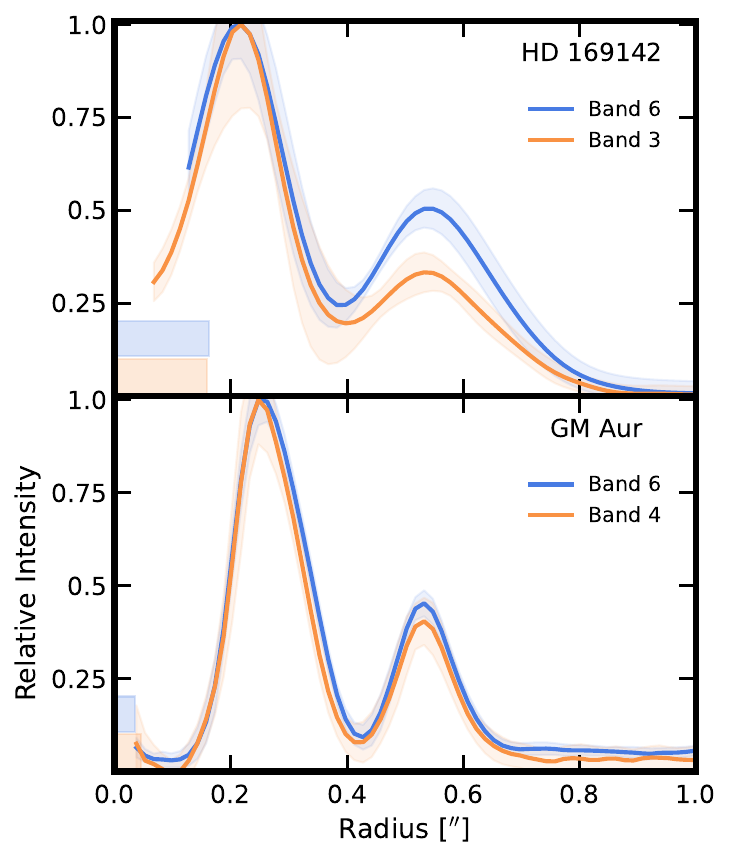}
    \caption{Azimuthally averaged relative intensity profiles of HD 169142 (upper panel) and GM Aur (lower panel) at Bands 6 (blue) and 3/4 (orange). The shaded regions show the relative standard deviation. The coloured bars at the left corner show the beam size of observations encoded in the same colour. The profiles are obtained after deprojecting targets on the image plane by their position angles and inclination \citep{2006AJ....131.2290R, 2020ApJ...891...48H}. The upper panel is created with data from \citet{2019ApJ...881..159M} (Band 3) and data retrieved from the ALMA science archive (Band 6). The resolution and sensitivity are $0.^{\prime\prime}22\times 0.^{\prime\prime}10$ (PA$=-88^{\circ}$) and $17~\mu\mathrm{Jy~beam}^{-1}$ for the Band 3 observation, and $0.^{\prime\prime}19\times 0.^{\prime\prime}13$ (PA$=63^{\circ}$) and $108~\mu\mathrm{Jy~beam}^{-1}$ for the Band 6 observation. The lower panel is created based on data from \citet{2020ApJ...891...48H}. The resolution and sensitivity are $0.^{\prime\prime}057\times 0.^{\prime\prime}034$ (PA$=-13^{\circ}$) and $12~\mu\mathrm{Jy~beam}^{-1}$ for the Band 4 observation, and $0.^{\prime\prime}045\times 0.^{\prime\prime}025$ (PA$=2^{\circ}$) and $10~\mu\mathrm{Jy~beam}^{-1}$ for the Band 6 observation.}
    \label{fig:brightness}
\end{figure}

Two common features of recent multi-wavelength Atacama Large Millimetre/submillimetre Array (ALMA) observations are: a. dust rings seen at short wavelengths (e.g. Band 6/7; $\lambda=1.3/0.89~\mathrm{mm}$) also appear at longer wavelengths (e.g. Band 3/4; $\lambda=3/2.1~\mathrm{mm}$); and b. the brightness ratio of multiple dust rings within a given disc remains largely consistent across wavelengths\footnote{For example, the brightness ratio of the inner to outer rings is 2.21 at Band 6 and 2.47 at Band 4 for GM Aur, and 1.98 at Band 6 and 3.00 at Band 3 for HD 164192. The brightness ratio therefore increases by factors of 1.12 and 1.51 when moving to longer wavelengths for GM Aur and HD 164192, respectively. The beam size is comparable for each target across wavelengths.}. Such trends have been found for a number of protoplanetary discs with ring-like substructures, such as HL Tau \citep{2015ApJ...808L...3A, 2019ApJ...883...71C, 2024A&A...686A.298G}, HD 163296 \citep{2018ApJ...869L..41A, 2023ApJ...957...11D}, HD 164192 \citep{2019ApJ...881..159M}, MWC 480 \citep[][Shi et al, in preparation]{2021ApJS..257....1O}, GM Aur \citep{2020ApJ...891...48H}, etc. Two representative examples, HD 164192 and GM Aur, are shown in Fig. \ref{fig:brightness}, demonstrating the two common features are not limited to a specific disc. Though other discs are not shown here, they follow the similar trend. The prevalence of the phenomenon also indicates that the two common features of dust rings should not be transient, but instead must persist for a large fraction of the disc lifetime. This adds an additional constraint on time-scales: that these two observational facts should remain true for at least a few Myr. These three observational facts can impose more direct and stringent constraints on the dust size distributions and the allowable combinations of dust fragility and disc turbulence levels.

In this section, we implement \dustpy models to test the combination(s) of \vfrag and $\alpha$ that can fulfil the three requirements discussed above with assistance of synthetic observations generated from radiative transfer. We do not aim to reproduce the exact values of brightness ratios of dust rings within a single wavelength and across multi-wavelengths, but require that the brightness ratio of the inner to outer dust rings are consistently $>1$ or $<1$ and that the ratio does not change significantly over a sufficiently long time ($\gtrsim$ Myr) and across wavelengths  instead. We extend our investigation to disc continuum fluxes at $\lambda=1.3~$ and $3~\mathrm{mm}$\footnote{We note that ALMA Band 3 ($\lambda=3~\mathrm{mm}$) is used instead of Band 4 ($\lambda=2.1~\mathrm{mm}$), as Band 3 observations probe continuum emission from larger grains and represent the longest wavelength commonly used in ALMA studies.}, disc-integrated spectral indices, optical depth and dust retention, i.e. how the dust mass changes over time. We adopt the disc and stellar parameters of GM Aur from Table \ref{tb:stellar_params}. Our goal is \textit{not} to reproduce the exact observables of GM Aur, but rather to explore combinations of \vfrag and $\alpha$ that can describe several observables within reasonable ranges found in recent observations of general discs. We describe our models in Section \ref{subsec:numerical_methods} and present our results in Section \ref{subsec:numerical_results}.

%% file: mainbody/numerical_methods.tex
\subsubsection{Gas evolution}
We model gas and dust evolution using \textsc{DustPy}. Gas evolution is driven by $\alpha$-parametrised viscosity \citep{1973A&A....24..337S} and follows the diffusion equation \citep{1974MNRAS.168..603L}:
\begin{equation}\label{eq:master}
    \frac{\partial \Sigma_g}{\partial t}= \frac{3}{R}\frac{\partial}{\partial R}\Biggl[R^{1/2}\frac{\partial }{\partial R
    }\bigg(\nu \Sigma_g R^{1/2}\bigg)\Biggr],
 \end{equation}
where $\Sigma_g$ is the gas surface density and has an initial profile of
\begin{equation}\label{eq:sigma_gas}
     \Sigma_g = \frac{M_d}{2\pi R_c^2}\bigg(\frac{R}{R_c}\bigg)^{-1}\exp{\bigg(-\frac{R}{R_c}\bigg)},
\end{equation}
depending on the disc mass $M_d$ and the characteristic radius $R_c$. $\nu=\alpha c_sH_g$ in Equation \ref{eq:master} is kinematic viscosity. $c_s=\sqrt{k_BT/(\mu m_H)}$ is the speed of sound and $H_g=c_s/\Omega_K$ is the gas scale-height. $k_B$ is the Boltzmann constant, and $T$ is the mid-plane temperature. $m_H$ is the mass of atomic hydrogen and $\mu=2.3$ is the mean molecular mass in units of $m_H$. $\Omega_K=\sqrt{GM_*/R^3}$ is the Keplerian angular velocity orbiting a central star of $1~M_*$ at the radius $R$. $G$ is the gravitational constant. The disc mid-plane temperature is consistently computed from stellar parameters by 
\begin{equation}\label{eq:thermal}
	T(R) = \bigg(\frac{\varphi L_*}{8\pi R^2\sigma_\mathrm{SB}}\bigg)^{1/4},
\end{equation}
where $L_*=4\pi R_*^2\sigma_\mathrm{SB}T_\mathrm{eff}^4$ is the stellar luminosity, depending on the stellar radius $R_*$ and the stellar effective temperature $T_\mathrm{eff}$. $\varphi$ is the irradiation angle and $\sigma_\mathrm{SB}$ is the Stefan-Boltzmann constant. Under viscous diffusion, gas has a radial velocity of 
\begin{equation}\label{eq:v_gas}
    v_g = -\frac{3}{\Sigma_g\sqrt{R}}\frac{\partial}{\partial R}\bigg(\Sigma_g \nu\sqrt{R}\bigg).
\end{equation}

\subsubsection{Dust evolution}
Dust evolution in \textsc{DustPy} accounts for dust transport and dust collisions. Dust evolution depends on gas evolution but its feedback on gas evolution is not considered by default in \textsc{DustPy}. Such treatment is good approximation as long as gas is the dominant material.

The coupling between gas and dust is described by Stokes number $\mathrm{St}$. Particles with sizes smaller than the gas mean free path $\lambda_\mathrm{mfp}$ ($a_i<9/4 \lambda_\mathrm{mfp}$) fall in the Epstein regime, where 
\begin{equation}\label{eq:stokes}
    \mathrm{St}= \frac{\pi}{2}\frac{a_i \rho_s}{\Sigma_g}.
\end{equation}
$\rho_s$ is the dust bulk density. Larger particles with sizes of $a_i>9/4\lambda_\mathrm{mfp}$ experience gas as a flow and fall in the Stokes regime. 

Dust transport is described by an advection-diffusion equation \citep{2010A&A...513A..79B}
\begin{equation}
    \frac{\partial \Sigma_d^i}{\partial t}+\frac{1}{R}\frac{\partial}{\partial R}\Big(R\Sigma_d^i v_{R,d}^i\Big)- \frac{1}{R}\frac{\partial}{\partial R}\Bigg[RD_d^i\Sigma_g \frac{\partial}{\partial R}\Bigg(\frac{\Sigma_d^i}{\Sigma_g}\Bigg)\Bigg]=0,
\end{equation}
where $\Sigma_d^{i}$ is the dust surface density of species $i$,  $v_{R,d}^{i}$ and $D_d^{i}$ are the radial advection velocity and dust diffusivity of species $i$, respectively. The cumulative dust surface density integrating over the dust species $\Sigma_d(R)$ is related to the gas surface density (Equation \ref{eq:sigma_gas}) by the dust-to-gas mass ratio $\epsilon(R)=  \Sigma_d(R) / \Sigma_g(R)$. Dust diffusivity describes the turbulent mixing for dust, specifically for the radial direction here, by
\begin{equation}
    D_d^{i} = \frac{\delta_\mathrm{rad}c_s^2}{\Omega_k(1+\mathrm{St_{i}}^2)},
\end{equation}
where $\delta_\mathrm{rad}$ is the radial turbulent mixing parameter, and it takes the value of gas turbulence $\alpha$ by default.
 
Dust orbiting central stars follows Keplerian motion in vacuum, while gas follows sub-Keplerian motion due to the presence of a negative radial pressure gradient. The drag force between gas and dust reduces their velocity difference and contributes to the dust radial drift velocity as 
\begin{equation}
    v_{R,\mathrm{drag}} = \frac{1}{1+\mathrm{St}_i^2}v_g,
\end{equation}
where $v_g$ is the gas radial velocity from diffusion (Equation \ref{eq:v_gas}). In addition to the gas drag, loss of angular momentum--arising from the ``headwind'' felt by dust from slower-moving gas--induces another radial velocity component
\begin{equation}
    v_{R,\mathrm{dust}}=-\frac{2\mathrm{St}_{i}}{1+\mathrm{St}_{i}^2}\eta v_K.
\end{equation}
$v_K = \Omega_K R$ is the Keplerian velocity and 
\begin{equation}
    \eta = -\frac{1}{2}\Bigg(\frac{H_g}{R}\Bigg)^2\frac{d \ln P}{d \ln R}
\end{equation}
denotes the deviation of gas motion from the Keplerian motion. $P=\rho_{g,\mathrm{mid}} c_s^2$ is the gas pressure and $\rho_{g,\mathrm{mid}}$ is the gas volume density in the mid-plane. The dust radial velocity is hence \citep{1986Icar...67..375N}
\begin{equation}
    v_{R,d}^{i} = v_{R,\mathrm{drag}}+v_{R, \mathrm{dust}} = \frac{1}{1+\mathrm{St}_i^2}v_g -\frac{2\mathrm{St}_{i}}{1+\mathrm{St}_{i}^2}\eta v_K.
\end{equation}

The dust size distribution after collisions are determined by solving the Smoluchowski equation \citep{1916ZPhy...17..557S}. Whether dust coagulates or fragments is determined by the relative velocity of two colliders, which accounts for the relative radial velocity, vertical settling, azimuthal velocity, turbulent motion and Brownian motion. Dust grows when the relative velocity is below a threshold, i.e. the fragmentation velocity, and fragments when it exceeds the threshold. The default setting in \textsc{DustPy} does not consider bouncing, where two colliders are scattered away with no mass loss or transfer. Bouncing is included in the model via the velocity prescription in \citet{2010A&A...513A..56G}, but we defer discussion of bouncing effects to Section \ref{subsec:discussion_dust_prop} (bouncing is switched off in the code until then). 

\subsubsection{Pressure bumps}\label{subsubsec:pressure}
Dust is lost rapidly without effective mechanisms to retain it within the disc. In addition to having massive gas discs or more porous dust to decrease the Stokes number, pressure bumps are commonly invoked to slow down dust radial drift and concentrate dust locally \citep[e.g.][]{1972fpp..conf..211W, 2012A&A...538A...114P}. Without addressing the origins of pressure bumps, we modify the gas surface density by using modified $\alpha$ profile to mimic the effects of pressure bumps as in \citet{2023A&A...679A..15G},
\begin{equation}\label{eq:bump}
	\alpha_\mathrm{pert} = \alpha \cdot \Bigg[1+ A_\mathrm{g}\cdot \exp \Bigg(-\frac{(R-R_\mathrm{g})^2}{2w_\mathrm{g}^2} \Bigg) \Bigg],
\end{equation}
where $A_\mathrm{g}$, $R_\mathrm{g}$, and $w_\mathrm{g}$ are the amplitude, median and width of the Gaussian function employed to modify the gas turbulence, respectively. We adopt $w_\mathrm{g}=2H_g$ to ensure the stability of the pressure bumps \citep[e.g.][]{2000ApJ...533.1023L}. Turbulent mixing factors, such as $\delta_\mathrm{rad}$, are not altered by this perturbation.

\subsubsection{Setup}\label{subsec:setup}
The radial grid contains 1000 cells logarithmically spaced between 5 and 1000 au. The particle grid depends on the fragmentation velocity \vfrag. When \vfrag$\leq 1~\mathrm{m~s^{-1}}$, the particle grid is $0.5~\mathrm{\mu m}$--$24.2~\mathrm{cm}$ over 120 cells. When \vfrag$> 1~\mathrm{m~s^{-1}}$, the particle grid is $0.5~\mathrm{\mu m}$--$242~\mathrm{cm}$ over 141 cells. The initial dust size distribution in a given cell follows $n(a)\propto a^{-3.5}$ \citep{1977ApJ...217..425M} with the initial maximum particle size capped by either $1~\mathrm{\mu m}$ or a "safe" size that does not induce rapid drift, whichever is smaller. We assume the minimum dust size in the particle grid as the dust monomer size $a_0=0.5~\mathrm{\mu m}$ and assume a dust bulk density $\rho_s=1.67~\mathrm{g~cm^{-3}}$. 

As the dust surface density $\Sigma_d^{i}$ depends on discretised dust sizes in \dustpy, a grid-independent vertically averaged surface density 
\begin{equation}
    \sigma_d(R) = \int_{-\infty}^{+\infty} n(R, z, a)\cdot m(a) \cdot a~dz
\end{equation}
is defined to eliminate the effects of discretisation \citep{2010A&A...513A..79B}. Here, $a$ denotes the particle size, $m(a) = 4/3\pi a^3 \rho_s$ is the mass of a particle of size $a$, and $n(R,z,a)$ is the number density of particle with size $a$ at ($R, z$).

We adopt disc and stellar parameters of GM Aur, which are shown in Table \ref{tb:stellar_params}. This includes the stellar mass $M_*=1.13~M_\odot$, the stellar luminosity $L_*=1.2~L_\odot$, the effective temperature $T_\mathrm{eff}=4350~\mathrm{K}$, the initial gas disc mass $M_g=0.118~M_\odot$ and the gas disc characteristic size $R_c=96~\mathrm{au}$. The stellar radius, which is an input parameter for \dustpy models, is computed from $L_*=4\pi R_*^2\sigma T_\mathrm{eff}^4$. The initial dust-to-gas mass ratio is set to be 0.01. Our exploration covers 2 turbulence levels $\alpha= 10^{-4}$ and $10^{-3}$, which are typical values that are assumed in 1-D models and hydrodynamic simulations, and 14 fragmentation velocities \vfrag, ranging from $0.5$ to $12~\mathrm{m~s^{-1}}$, a common range that has been widely used in previous dust evolution models. All of these parameters are summarised in Table \ref{tb:params}.

\begin{figure}
    \centering
    \includegraphics[width=0.99\linewidth]{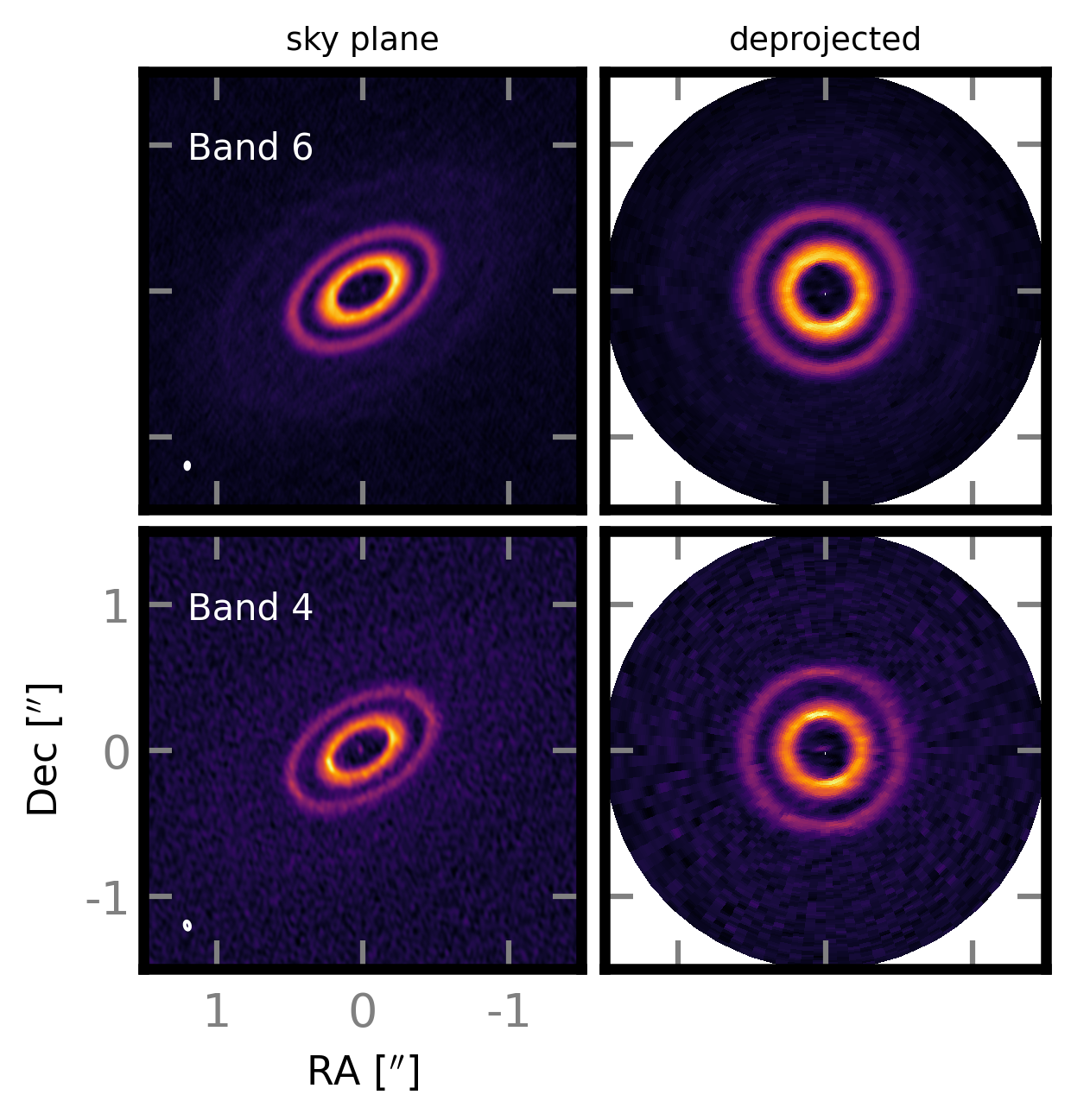}
    \caption{Continuum images of GM Aur in ALMA Bands 6 ($\lambda=1.3$ mm, upper panels) and 4 ($\lambda=2.1$ mm, lower panels). The left columns are for sky-plane images adopted from \citet{2020ApJ...891...48H}, and the right columns are for corresponding deprojected images. Beam sizes are denoted as ellipses at the left corners in panels of the sky plane images. Resolution and sensitivity of these data are reported in the caption of Fig. \ref{fig:brightness}. We remind the readers that GM Aur is shown as a representative example of multiple-ringed discs seen in multi-wavelength observations, and reproducing the GM Aur disc is not the aim of this study.}
    \label{fig:gmaur}
\end{figure}

We include two pressure bumps located at $R_\mathrm{g,1}=35~\mathrm{au}$ and $R_\mathrm{g,2}=70~\mathrm{au}$, with $A_{g,1}=3$ and $A_{g,2}=1.5$, respectively (see Equation \ref{eq:bump}). These results in two dust rings at approximately 45 and 90 au, mimicking the two rings in GM Aur (images shown in Fig. \ref{fig:gmaur}). The two pressure bumps are introduced simultaneously, and have reached stable amplitudes from the start of simulations. This is achieved by pre-evolving the gas-only disc to a relaxed state as described in \citet{2025MNRAS.537.3525T}. We pre-evolve discs with $\alpha=10^{-3}$ and $10^{-4}$  for 1 Myr and 1.5 Myr, respectively, as the latter has a longer viscous timescale. We explore the appearance order of two pressure bumps and their separations in Section \ref{subsec:discussion_dust_rings}.

Each model is evolved for 3 Myr after introducing the dust, i.e. additional 3 Myr after the pre-evolution. Disc properties and observables are read at $t=1$ and $3$ Myr\footnote{The gas disc mass decreases over time and becomes slightly lower than the gas disc mass of GM Aur measured today, which we assume as the initial gas disc mass. For models with $\alpha=10^{-4}$, it falls to $93.67$ per cent of the initial gas mass by 1 Myr and $92.52$ per cent by 3 Myr; for models with $\alpha=10^{-3}$, the corresponding fractions are $84.83$ per cent and $77.44$ per cent, respectively.}. We report the detectability of dust rings in synthetic observations, continuum fluxes measured from mock observations at Bands 3 ($\lambda=3~\mathrm{mm}$) and 6 ($\lambda=1.3~\mathrm{mm}$), as well as disc-integrated spectral indices derived from them in Table \ref{tb:model_flux}. A model is considered as valid when it meets criteria below:
\begin{itemize}
    \item Both the inner and outer dust rings are visible at Bands 3 and 6 over the evolutionary period 1--3 Myr;
    \item the brightness ratio of the inner to outer rings does not change significantly across two wavelengths and over time; and
    \item the disc-integrated spectral index $\alpha_\mathrm{1.3-3mm} = \mathrm{d}\log F_\mathrm{\nu}/ \mathrm{d}\log\mathrm{\nu}$
    falls in the typical ranges reported in previous studies \citep[$2\lesssim\alpha_\mathrm{1.3-3mm} \lesssim 4$, ][]{2019ApJ...883...71C, 2021MNRAS.506.5117T, 2025A&A...694A.290G}.
\end{itemize}

\begin{table}
\caption{Input parameters for our disc models.}\label{tb:params}
\begin{tabular}{ccc}
\hline
\hline
Parameters & Symbols [units] & Values\\   
\hline
Stellar mass  & $M_*$ [$M_\odot$] & 1.13 \\
Stellar radius & $R_*$ [$R_\odot$]  & 1.79 \\
Stellar luminosity & $L_*$ [$M_\odot$] & 1.2 \\
Effective stellar temperature   & $T_\mathrm{eff}$ [$\mathrm{K}$] & 4350\\
Initial gas disc mass & $M_g$ [$M_\odot$]  & 0.118\\
Initial dust-to-gas mass ratio & $\epsilon$  & 0.01 \\
Characteristic radius $^{*}$& $R_c$ [$\mathrm{au}$] & 96*\\
Turbulence level  & $\alpha$ & $10^{-4}$, $10^{-3}$\\
Fragmentation velocity  & $v_\mathrm{frag}$  [$\mathrm{m~s^{-1}}$] & 0.5, 1, 2, 4, 6, 8, 10, 12 \\
Dust monomer size &  $a_0 [\mathrm{\mu m}]$ & 0.5 \\ 
Dust bulk density & $\rho_s$ [$\mathrm{g~cm^{-3}}$]  & 1.67 \\ 
Distance to the disc & $d$ [$\mathrm{pc}$] & 150\\
Irradiation angle & $\varphi$ & 0.05\\
\hline
\end{tabular}
\footnotesize{\textbf{Note}: $^{*}$The characteristic radius shown here is the radius before the gas surface density is relaxed.}\\
\end{table}

\subsubsection{Radiative transfer and synthetic observations}\label{subsec:RT}
\dustpy models are used to generate synthetic observations for comparison with observations. This is achieved by post-processing radiative transfer models through the interface built in \textsc{DUSTPYLIB} \citep{2023ascl.soft10005S} to \textsc{RADMC-3D} \citep{2012ascl.soft02015D}\footnote{\url{https://www.ita.uni-heidelberg.de/~dullemond/software/radmc-3d/index.php}} .

The radiative transfer is implemented in spherical coordinates, with $10^{7}$ and $10^{6}$ photons for thermal radiation and anisotropic scattering, respectively. The radial grid is identical to that in \dustpy (see Section \ref{subsec:setup}). The azimuthal direction includes 16 cells from 0 to $2\pi$ and the polar direction includes 256 cells from 0 to $\pi$. The minimum value of the particle grid in \textsc{RAMDC-3D} is determined by the minimum dust size in the \dustpy models. If the maximum dust size in \dustpy models is $<10~\mathrm{cm}$, we use 50 logarithmically spaced particle bins up to $10~\mathrm{cm}$. If the maximum dust size $>10~\mathrm{cm}$, we use 80 logarithmically spaced bins extending up to $100~\mathrm{cm}$ instead.

The disc is assumed to be vertically isothermal in radiative transfer and the dust volume density $\rho_{d}^i$ of species $i$ follows 
\begin{equation}\label{eq:volume_density}
    \rho_{d}^i(z) = \frac{\Sigma_{d}^i}{\sqrt{2\pi}H_d^i}\exp \Bigg(-\frac{z^2}{{H^i_d}^2}\Bigg),
\end{equation}
where the dust scale height is \citep{1995Icar..114..237D}
\begin{equation}\label{eq:dust_scaleheight}
    H_{d}^i=H_g\sqrt{\frac{\delta_\mathrm{vert}}{\mathrm{St}_i+\delta_\mathrm{vert}}}.
\end{equation}
$\delta_\mathrm{vert}$ is the vertical turbulence mixing parameter and is assumed to be equal to $\alpha$ by default.

We adopt DSHARP opacities \citep{2018ApJ...869L..45B}\footnote{\url{https://github.com/birnstiel/dsharp opac/}} in radiative transfer models. This assumes compact dust composed of 2.58 per cent troilite, 16.7 per cent astronomical silicates \citep{2003ARA&A..41..241D}, 36.42 per cent water ice \citep{2008JGRD..11314220W} and 44.30 per cent refractory organics \citep{1996A&A...311..291H}. The effect introduced by opacities is discussed in Section \ref{subsec:opacity}, where we also consider the opacities from \citet{2010A&A...512A..15R}. This opacity follows the assumptions on the proportion in \citet{1994ApJ...421..615P} and assumes $30$ per cent porous dust composed of $7$ per cent astronomical silicates \citep{2001ApJ...548..296W}, $21$ per cent carbonaceous material \citep{1996MNRAS.282.1321Z} and $42$ per cent water ice \citep{1984ApOpt..23.1206W}. All percentages above are given as volume fractions.

The radiative transfer models are further employed to generate synthetic observations via \textsc{CASA} \citep{2022PASP..134k4501C}. We assume an on-source integration time of 30 mins with resolution of 0.05 arcsec. This leads to average sensitivity of 17.72~$\mu$Jy/beam in Band 6 and 15.93~$\mu$Jy/beam in Band 3, as derived from the valid models (Section \ref{subsubsec:valid_models}). The resolution and sensitivity of synthetic observations are comparable to those of DSHARP observations \citep{2018ApJ...869L..41A}. The synthetic observations are produced for face-on sources (inclination angle $i=0^{\circ}$) at ALMA Bands 3 ($\lambda=3~\mathrm{mm}$) and 6 ($\lambda=1.3~\mathrm{mm}$) with thermal noise considered. 

%% file: mainbody/numerical_results.tex
\subsubsection{Valid models}\label{subsubsec:valid_models}
\begin{figure}
    \centering
    \includegraphics[width=0.99\linewidth]{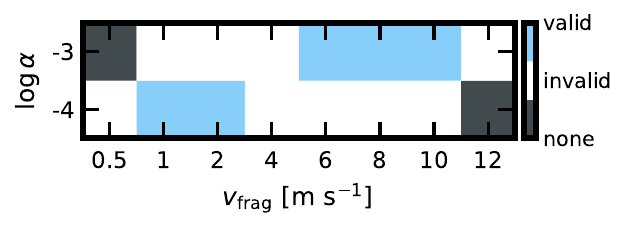}
    \caption{Overview of numerical models: blue indicates valid combinations that can fulfil the criteria in Section \ref{subsec:setup}; white indicates invalid models and grey indicates combinations that have not been tested due to extreme values of parameters.}
    \label{fig:valid_table}
\end{figure}

\begin{figure*}
    \centering
    \includegraphics[width=0.95\linewidth]{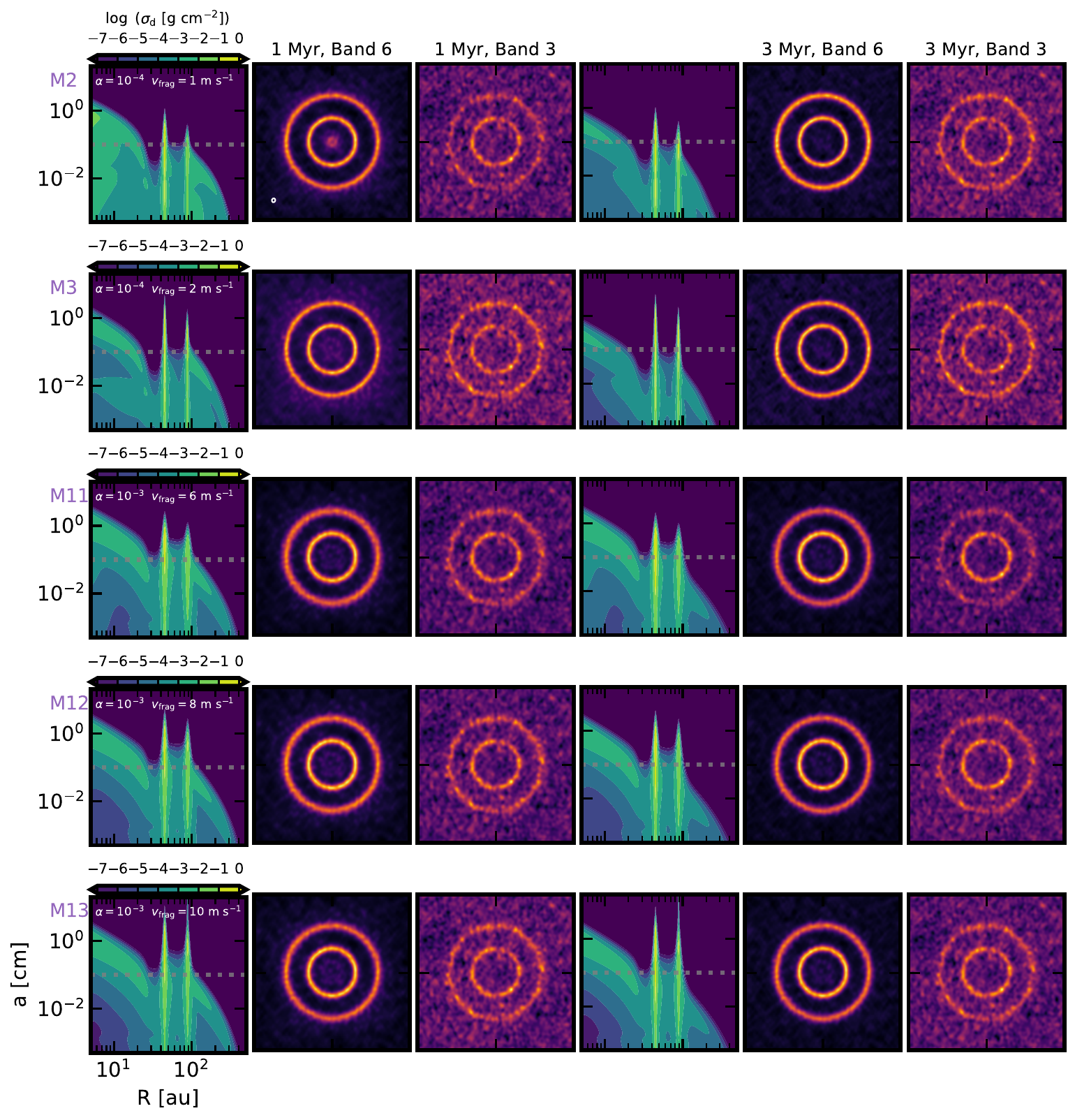}
    \caption{Valid numerical models (columns 1 and 4) and corresponding synthetic observations (columns 2-3 and 5-6). For each model, Bands 6 and 3 continuum synthetic observations are plotted at $t=1$ and $3~\mathrm{Myr}$. A representative beam size ($0.053$ arcsec $\times$ 0.045 arcsec, $15.^{\circ}28$) for the synthetic observations is indicated in the left lower corner of the panel in row 1, column 2. Synthetic images are plotted in the linear colour scale. The dotted lines plotted over the numerical models indicate the grain size $a=0.1~\mathrm{cm}$.}
    \label{fig:valid_model}
\end{figure*}

We run 14 \dustpy models with combinations of $\alpha=10^{-4}$/$10^{-3}$ and \vfrag$=0.5$--$12~\mathrm{m~s^{-1}}$ and show the result in Table \ref{tb:model_flux} and Fig. \ref{fig:valid_table}. The results support the analytical findings in Section \ref{sec:analytical} that only appropriate combinations of disc turbulence and dust fragility can reproduce the general trend in multi-wavelength observations. We present dust size distributions for numerical models at $t=1$ and $3$ Myr as well as their synthetic observations at $\lambda=1.3$ and $3.0~\mathrm{mm}$ in Fig. \ref{fig:valid_model} for valid models and in Fig. \ref{fig:invalid_model} for invalid models. We remind readers that the continuum images of GM Aur shown in Fig. \ref{fig:gmaur} are only for illustrative purposes and that reproducing them is not the aim of this study.

When turbulence is low ($\alpha=10^{-4}$), fragile dust with \vfrag of $1$--$2~\mathrm{m~s^{-1}}$ is required to reproduce two dust rings consistently at Bands 3 and 6 over the evolution. Under this given turbulence, more fragile dust with \vfrag$<1~\mathrm{m~s^{-1}}$ struggles to grow to larger sizes that can be trapped around the pressure bump in the outer disc (see the first row in Fig. \ref{fig:invalid_model}). These small particles have lower opacities at millimetre observing wavelengths, rendering a nearly invisible outer dust rings at $\lambda=3~\mathrm{mm}$ (Band 3). When dust becomes more resilient (\vfrag $\gtrsim 4~\mathrm{m~s^{-1}}$), it continues to grow into centimetre regime and concentrates most dust mass to large particles, as shown in Models 5-7 (see Fig. \ref{fig:invalid_model}). The opacity of these large particles is also low in millimetre wavelengths, again resulting discs barely visible in ALMA synthetic observations. 

Similarly, more turbulent discs ($\alpha=10^{-3}$) require more resilient dust (\vfrag $=6$--$10~\mathrm{m~s^{-1}}$). But too resilient dust, like the one with \vfrag$>10\mathrm{m~s^{-1}}$ (Model 14 in Fig. \ref{fig:invalid_model}), again clusters in particles with sizes $a>10$ cm and fails to have two dust rings across two wavelengths through the 3-Myr evolution. Though infinitely increasing sensitivity can enable the detectability of some faint dust rings, we remind that any dust rings invisible in the current sensitivity are not the dust rings commonly seen in recent observations.

In the following analysis in Section \ref{subsec:numerical_results} and discussion in Section \ref{sec:discussion}, we mainly focus on the five valid models. We show in Section \ref{subsec:result_intensity_ratio} and \ref{subsec:result_index} how the other two criteria listed in Section \ref{subsec:setup} are fulfilled by these 5 valid models. We briefly analyse invalid models in Appendix \ref{sec:appendix_invalid}.

\begin{table*}
\caption{Observable properties of Models 1-14 with various combinations of $\alpha$ and \vfrag. Column 2 is for the turbulence level and column 3 is for the fragmentation velocity \vfrag. Columns 4-7 show the detectability of the inner and outer dust rings in each model: \cmark~for visible, \xmark~for invisible, and \bqm~for tentatively visible (tentatively visible in image but has an S/N<3). Two markers are shown in each column, with the first and second symbols representing the result for the inner and outer dust rings, respectively. Columns 8-11 shows the continuum fluxes measured at the indicated time and wavelengths from synthetic observations. Columns 12-13 shows the spectral indices inferred from columns 8-11. Entries are left blank if the disc is not detected at either wavelength. In Column 14 we use \cmark~to indicate models that meet the criteria of valid models described at the end of Section \ref{subsec:setup}}. \label{tb:model_flux}

\begin{tabular}{cccccccccccccc}
\hline
\hline
(1) & (2) & (3) & (4) & (5) & (6) & (7) & (8) & (9) & (10) & (11) & (12) & (13) & (14) \\
Model &  $\alpha$& $v_\mathrm{frag}$ & $R_\mathrm{1.3mm}^\mathrm{1myr}$ & $R_\mathrm{3mm}^\mathrm{1myr}$ &  
$R_\mathrm{1.3mm}^\mathrm{3myr}$ & $R_\mathrm{3mm}^\mathrm{3myr}$ & 
$F_\mathrm{1.3mm}^\mathrm{1myr}$ &$F_\mathrm{3mm}^\mathrm{1myr}$ &$F_\mathrm{1.3mm}^\mathrm{3myr}$ & $F_\mathrm{3.0mm}^\mathrm{3myr}$ & $\alpha_\mathrm{1.3-3mm}^\mathrm{1myr}$ & $\alpha_\mathrm{1.3-3mm}^\mathrm{3myr}$ & valid model? \\
 &  & [$\mathrm{m~s^{-1}}$] & in/out & in/out  & in/out & in/out & [mJy] & [mJy] & [mJy] & [mJy] & & & \\
\hline
1 & $10^{-4}$ & $0.5$ &\cmark\cmark &\cmark\xmark &\cmark\cmark  &\cmark\bqm  & 156.96 & 10.39 & 120.78 & 7.75 & 3.25 & 3.28 & \\
2 & $10^{-4}$ & $1$ & \cmark\cmark & \cmark\cmark & \cmark\cmark & \cmark\cmark & 123.94 & 9.67 & 102.84 & 11.59 & 3.05 & 2.61 & \cmark \\
3 & $10^{-4}$ & $2$ & \cmark\cmark & \cmark\cmark & \cmark\cmark & \cmark\cmark & 99.66 & 9.73 & 76.13 & 9.50 & 2.78 & 2.49 & \cmark \\
4 & $10^{-4}$ & $4$ & \cmark\cmark & \cmark\bqm & \cmark\cmark & \cmark\xmark & 53.20 & 5.43 & 31.04 & 3.85 & 2.73 & 2.50 & \\
5 & $10^{-4}$ & $6$ & \cmark\cmark & \xmark\xmark & \bqm\bqm & \xmark\xmark & 17.66 & 1.58 & 4.64 & 0.56 & -- & -- &  \\
6 & $10^{-4}$ & $8$ & \bqm\cmark & \xmark\xmark & \xmark\xmark & \xmark\xmark & 15.75 & 1.32 & 3.29 & 0.34 & -- & -- & \\
7 & $10^{-4}$ & $10$& \bqm\cmark & \xmark\xmark & \xmark\xmark & \xmark\xmark & 15.02 & 1.29 & 3.21 & 0.33 & -- & -- & \\
8 & $10^{-3}$ & $1$& \cmark\xmark & \xmark\xmark & \cmark\xmark & \xmark\xmark & 129.10 & 5.82 & 109.68 & 3.30 & -- & -- & \\
9 & $10^{-3}$ & $2$& \cmark\cmark & \cmark\xmark & \cmark\cmark & \cmark\xmark & 225.25 & 7.95 & 198.94 & 7.84 & 3.99 & 3.87 & \\
10 & $10^{-3}$ & $4$& \cmark\cmark & \cmark\cmark & \cmark\cmark & \cmark\bqm & 199.12 & 15.62 & 153.73 & 12.56 & 3.04 & 3.00 & \\
11 & $10^{-3}$ & $6$& \cmark\cmark & \cmark\cmark & \cmark\cmark & \cmark\cmark & 166.21 & 15.08 & 138.18 & 13.22 & 2.87 & 2.81 & \cmark \\
12 & $10^{-3}$ & $8$& \cmark\cmark & \cmark\cmark & \cmark\cmark & \cmark\cmark & 149.07 & 13.48 & 126.45 & 13.40 & 2.87 & 2.68 & \cmark\\
13 & $10^{-3}$ & $10$& \cmark\cmark & \cmark\cmark & \cmark\cmark & \cmark\cmark &  139.96 & 13.55 & 118.66 & 12.71 & 2.79 & 2.67 & \cmark \\
14 & $10^{-3}$ & $12$& \cmark\cmark & \cmark\bqm & \cmark\cmark & \cmark\bqm &  89.09 & 8.94 & 53.15 & 6.37 & 2.75 & 2.54 & \\
\hline

\end{tabular}
\end{table*}

\subsubsection{Optical depth}\label{subsubsec:optical depth}
\begin{figure}
    \centering
    \includegraphics[width=0.99\linewidth]{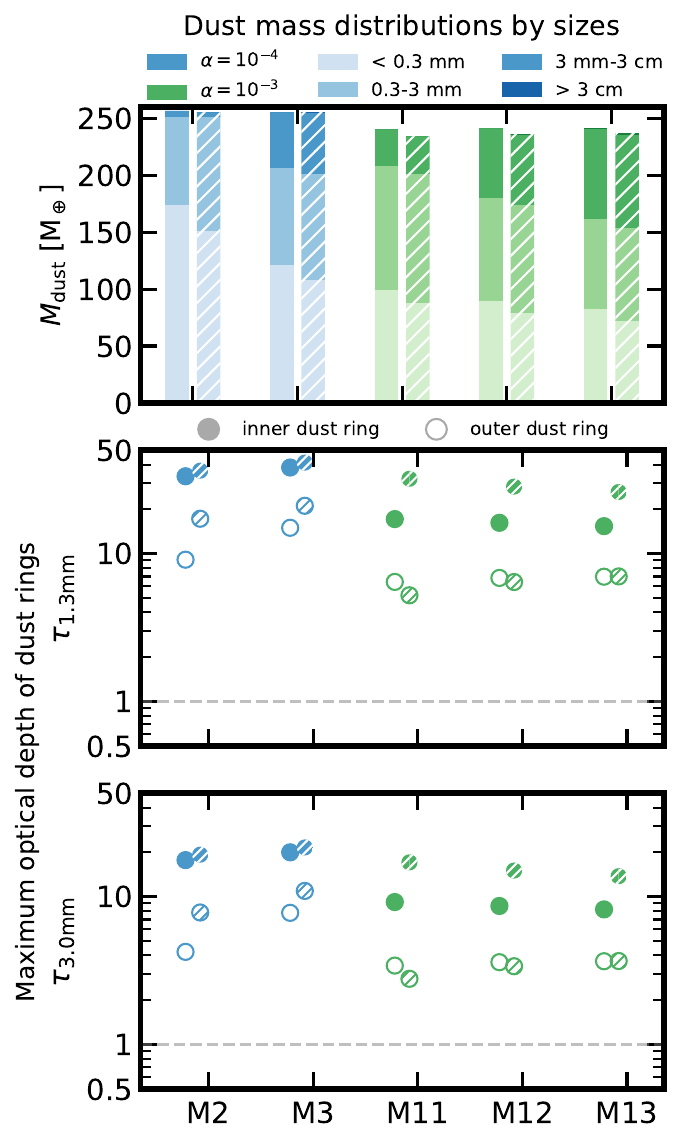}
    \caption{Dust size distributions and optical depth for five valid models shown in Fig \ref{fig:valid_model} at $1$ Myr (solid) and $3$ Myr (hatched). Upper panel: the height of coloured bars show the theoretical total dust mass. Each segment in a given bar shows the mass fractions of particles in the indicated size bin. The colour is encoded with turbulence strengths $\alpha$ (blue for $\alpha=10^{-4}$ and green for $\alpha=10^{-3}$). The transparency of bars is encoded with dust sizes, with lighter shades indicating smaller dust. Middle and lower panels: the maximum optical depth measured around the inner (filled dots) and outer (empty dots) dust rings at $\lambda=1.3$ mm (middle) and $\lambda=3.0$ mm (lower). The colours and filled patterns are the same as the upper panel. The dashed grey lines indicate where optical depth $\tau_\lambda=1$.}
    \label{fig:optical_depth}
\end{figure}

We measure the peak optical depth around the dust ring by 
\begin{equation}\label{eq:optical_depth}
    \tau_\nu = \sum_i \big(\kappa^\mathrm{sca,eff}_\mathrm{\nu,i}+\kappa^\mathrm{abs}_\mathrm{\nu,i}\big)\Sigma_{d}^{i},
\end{equation}
where $\kappa^\mathrm{abs}_\mathrm{\nu,i}$ and $\kappa^\mathrm{sca,eff}_\mathrm{\nu,i}$ are the absorption and effective scattering opacities of dust species $i$ at the frequency $\nu$; $\Sigma_{d}^{i}$ is the dust surface density of dust species $i$. The effective scattering opacity is prescribed by $\kappa_{\nu,i}^\mathrm{sca,eff}=(1-g_{\nu,i}) \kappa_{\nu,i}^\mathrm{sca}$, where $g_{\nu,i}$ is a parameter to correct the isotropic scattering. The values of $\kappa^\mathrm{abs}_\mathrm{\nu,i}$, $\kappa^\mathrm{sca}_\mathrm{\nu,i}$ and $g_{\nu,i}$ are adopted directly from \textsc{DSHARP} opacities \citep{2018ApJ...869L..45B}. The optical depth computed from valid numerical models is shown in the middle and lower panels of Fig. \ref{fig:optical_depth}.

Dust rings in all the 5 valid models are optically thick at 1.3 and 3 mm throughout the evolution, and the inner dust rings have higher optical depth than the outer dust rings for a given model. The inner dust rings serving as a more abundant dust reservoir is due to two reasons: a) inner discs are originally denser in dust; and b) dust which is not trapped by the outer pressure bumps drifts inwards and can be trapped around the inner pressure bumps.

Among the five valid models, those with lower turbulence ($\alpha=10^{-4}$) have higher optical depth ($\tau_\mathrm{1.3mm}$ and $\tau_\mathrm{3.0mm}$) for both dust rings. This is because more dust mass is concentrated in smaller particles ($<3$ mm), which contribute more to dust opacities at millimetre wavelengths in a less turbulent disc (see the upper panel of Fig. \ref{fig:optical_depth}).

The optical depth ($\tau_\mathrm{1.3mm}$ and $\tau_\mathrm{3.0mm}$) for dust rings increases after Myr-evolution, except for the outer dust rings in models with $\alpha$= $10^{-3}$, whose optical depth remains nearly the same. This increase reflects a shift in dust size distribution from $<0.3$ mm to $0.3$--$3$ mm. Since small particles that are not trapped by dust rings locally are lost from discs, this results in higher proportions of particles in $0.3$-$3$ mm, which contribute more to the opacity at $1.3$ and $3$ mm and increases the optical depth.

These trends are shared between $\lambda=1.3$ and $3.0$ mm. Though the optical depth for both dust rings is lower at longer wavelengths ($\tau_\mathrm{3.0mm}<\tau_\mathrm{1.3mm}$) as expected, it remains large ($>1$) over the entire evolution. Compared with the invalid models in Appendix \ref{sec:appendix_invalid}, being optically thick for both dust rings at both wavelengths across 3 Myr is required to reproduce the general trend of multi-wavelength observations. We note this result is derived from comparison between valid and invalid models presented in this paper rather than assumptions on the optical depth.

\subsubsection{Intensity ratios of dust rings}\label{subsec:result_intensity_ratio}

In the optically thick limit ($\tau_\nu\gg 1$), the intensity ratio of the inner ($R_1$) to the outer ($R_2$) dust rings can be written as a function of local temperature ratios in the Rayleigh-Jeans regime.
\begin{equation}\label{eq:intensity}
    \frac{I_\mathrm{\nu, R_1}}{I_\mathrm{\nu, R_2}}=\frac{(1-e^{-\tau_{\nu,1}})B_\nu (T_\mathrm{R_1})}{(1-e^{-\tau_{\nu,2}})B_\nu (T_\mathrm{R_2})}\simeq\frac{B_\nu (T_\mathrm{R_1})}{B_\nu(T_\mathrm{R_2})}=\frac{T_\mathrm{R_1}}{T_\mathrm{R_2}}.
\end{equation} 
Substituting Equation \ref{eq:thermal} into Equation \ref{eq:intensity} gives 
\begin{equation}
    \frac{I_\mathrm{\nu, R_1}}{I_\mathrm{\nu, R_2}}=\bigg(\frac{R_2}{R_1}\bigg)^{1/2}=\bigg(\frac{90}{45}\bigg)^{1/2}\sim 1.4.
\end{equation}
In the last two steps, we insert the radii of two dust rings of our models $R_1=45~\mathrm{au}$, and $R_2=90~\mathrm{au}$. This results in an intensity ratio of $\sim 1.4$.

This is a simplified estimate. In a more realistic case, we have to account for the dust vertical distribution and thermal stratification, yielding an intensity ratio which deviates from 1.4. We measure the intensity ratio of the inner to outer dust rings for valid models from synthetic observations in Fig. \ref{fig:valid_model} and summarized them in Table \ref{tb:model_int}. 

At a given time, the intensity ratio is not expected to vary significantly across observing wavelengths but slightly increases toward longer wavelengths (see Fig. \ref{fig:brightness}). This trend is generally reproduced by the valid models. Such a trend suggests that the inner rings become relatively brighter at longer wavelengths. We attributed this to the different dust size distributions between the inner and the outer dust rings. In the outer disc, smaller particles can achieve Stokes numbers comparable to those of larger ones in the inner disc due to its lower local gas surface density. This shifts the dust size distribution in the outer dust ring towards smaller sizes. These smaller particles have lower opacities at longer wavelengths, leading to weaker emission from the outer dust ring compared to that from the inner dust ring. 

In addition to the dust size distributions, the intensity ratio of the two dust rings also depends on the amplitude of pressure bumps, the separation of two dust rings and appearance orders of dust rings. The effects from the latter two will be discussed in Section \ref{subsec:discussion_dust_rings}. For the pressure bump models (Section \ref{subsubsec:pressure}), we examine another pair of pressure bump amplitudes $A_\mathrm{g,1}=2$ and $A_\mathrm{g,2}=3$ while keep other parameters the same as in Section \ref{subsec:setup}. The intensity ratios of dust rings of some models with new setup flip during the evolution--the outer ring is brighter at 1 Myr while the inner dust ring becomes brighter at 3 Myr. This is because the outer dust ring with a larger pressure gradient is more capable of retaining dust than the inner dust ring initially. However, the valid combinations of $\alpha$ and \vfrag that can reproduce multi-wavelength observations do not change with the pressure bump parameters explored here. 

\begin{table}
\caption{Intensity ratios of the inner to outer dust rings for valid models identified in Table \ref{tb:model_flux}.} \label{tb:model_int}
\begin{tabular}{ccccc}
\hline
\hline
(1) & (2) & (3) & (4) & (5)\\
Model &  $\mathrm{I_1/I_2}_\mathrm{1.3mm}^\mathrm{1myr}$ & $\mathrm{I_1/I_2}_\mathrm{3mm}^\mathrm{1myr}$ & $\mathrm{I_1/I_2}_\mathrm{1.3mm}^\mathrm{3myr}$ &  
$\mathrm{I_1/I_2}_\mathrm{3mm}^\mathrm{3myr}$\\
\hline
2 & 1.19 & 1.69 & 1.08 & 1.18 \\
3 & 1.15 & 1.01 & 1.09 & 0.96 \\
11& 1.34 & 1.54 & 1.48 & 2.24 \\
12& 1.30 & 1.34 & 1.38 & 1.65 \\
13& 1.31 & 1.28 & 1.34 & 1.47 \\
\hline
\end{tabular}
\end{table}

\subsubsection{Disc-integrated continuum fluxes and spectral indices}\label{subsec:result_index}

\begin{figure}
    \centering
    \includegraphics[width=0.95\linewidth]{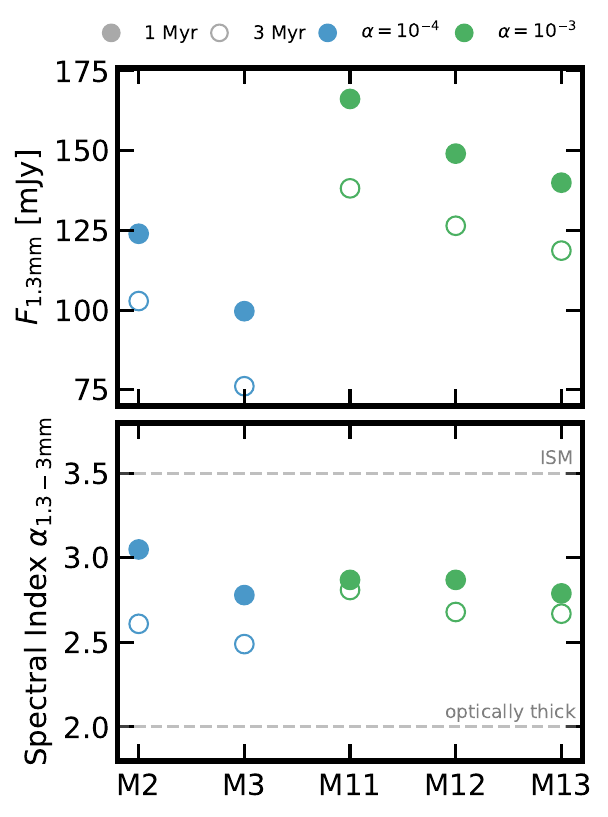}
    \caption{The continuum fluxes $F_\mathrm{1.3mm}$ (the upper panel) and the disc-integrated spectral indices between $\lambda=1.3~\mathrm{mm}$ and $3~\mathrm{mm}$ (the lower panel) measured for five valid models. Filled and empty dots are for measurements at 1 and 3 Myr, respectively. Colours encode the disc turbulence, with blue for $\alpha=10^{-4}$ and green for $\alpha=10^{-3}$.}
    \label{fig:flux_specral_ind}
\end{figure}

The disc continuum fluxes $F_\mathrm{1.3mm}$ and $F_\mathrm{3mm}$, and the disc-integrated spectral indices derived from them, are summarised in Table \ref{tb:model_flux} and are visualised in Fig. \ref{fig:flux_specral_ind}.

Even when the theoretical dust mass is similar (the upper panel of Fig. \ref{fig:optical_depth}), low-turbulence discs have $25$--$50$ per cent lower $F_\mathrm{1.3mm}$ than their high-turbulence counterparts because their emission is more optically thick (see the middle and lower panels of Fig. \ref{fig:optical_depth}). When the dust disc mass is derived under the standard optically thin assumption, lower-turbulence discs with fragile dust therefore yield an even lower dust disc mass than the true dust reservoir. Consequently, these weakly turbulent discs can more effectively mitigate the dust mass budget problem--the mass derived from millimetre continuum appears insufficient to reproduce exoplanet demographics \citep[e.g.][]{2014MNRAS.445.3315N, 2018A&A...618L...3M}.

For a given turbulence level, discs with lower \vfrag are brighter in continuum fluxes $F_\mathrm{1.3mm}$ over the entire evolution, even though the dust mass is nearly identical (the upper panel of Fig. \ref{fig:optical_depth}). As the total fluxes are dominated by two dust rings, this implies the brightness of optically thick dust rings varies with the dust size distribution set by \vfrag. In other words, despite being optically thick, the brightness of dust rings still depends on the dust size distribution, through its effects on the dust vertical distribution and the height of emitting layers, which introduce a different dust size distribution from the vertically averaged one at a given radius. This seems counter-intuitive to our typical expectation that the brightness of optically thick discs is independent of the dust density, an assumption which ignores the dust vertical distribution by adopting a vertically averaged surface density. 

The spectral indices are similar for all 5 valid models, from 2.5 to 3.0, a typical range for discs with substructures \citep[e.g.][]{2021MNRAS.506.5117T, 2025A&A...694A.290G}. We identify a decreasing trend in the spectral index over time, which we attribute to the loss of smaller particles ($<0.3~\mathrm{mm}$, see the upper panel of Fig. \ref{fig:optical_depth}). As particles with sizes similar to the observing wavelengths contribute more to the opacity, a shift in dust size distributions to larger sizes, as shown in Section \ref{subsubsec:optical depth}, induces larger optical depth at the two observing wavelengths (see the middle and lower panels of Fig. \ref{fig:optical_depth}) and flattens the differences between $F_\mathrm{1.3mm}$ and $F_\mathrm{3mm}$, leading to a lower spectral index. This may explain the slightly lower spectral indices found for discs in Upper Sco, compared to those in Lupus \citep{2025arXiv250706298Q}. However, we note that discs presented in \citet{2025arXiv250706298Q} have lower spectral indices than those shown here, and this could be caused by a variety of physics, such as the real opacities differing from the one assumed here (see Section \ref{subsec:opacity} and Table \ref{tb:model_flux_ricci10}, where the \citet{2010A&A...512A..15R} opacity is adopted and the spectral indices are generally lower), contamination by non-thermal emission \citep[e.g.][]{1975A&A....39....1P, 1986ApJ...304..713R, 2024A&A...684A.134R}, and contributions from circumstellar discs around binary systems, which typically have spectral indices $<2$ \citep{2021MNRAS.507.2531Z}. 

%% file: mainbody/discussion.tex
\subsection{Why \vfrag increases with dust ring radius}\label{subsec:discussion_increasing_vfrag}

Fig. \ref{fig:ana_sol_vfrag} in Section \ref{sec:analytical} shows that the fragmentation velocity is not only a function of turbulence, but also an increasing function of the radius of the dust rings when the turbulence is homogeneous and fixed. In this section, we discuss the underlying physics of this phenomenon, starting from analysing the optical depth and thermal emitting layer.

\begin{figure}
    \centering
    \includegraphics[width=0.85\linewidth]{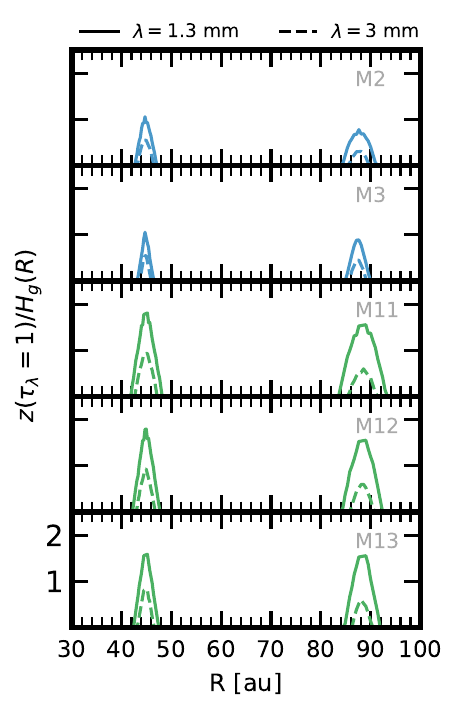}
    \caption{The emitting layer of thermal emission $\tau_\mathrm{1.3mm}=1$ (solid lines) and $\tau_\mathrm{3mm}=1$ (dashed lines) in units of gas scale-height $H_g(R)$ for five valid models: Models 2-3 ($\alpha=10^{-4}$, the upper two panels) and Models 11-13 ($\alpha=10^{-3}$, the lower three panels). All the models are at $t=1~\mathrm{Myr}$}.
    \label{fig:emit_layer}
\end{figure}

As shown in Section \ref{subsubsec:optical depth}, the optical depth at $\lambda=1.3$ and $3.0~\mathrm{mm}$ around dust rings for all valid models is greater than unity in the first 3 Myr (see the middle and lower panels of Fig. \ref{fig:optical_depth}). This suggests the emission we observe is not from the mid-plane but from the upper layer where $\tau_\lambda\sim1$. To trace the layer where the observed emission is from ($\tau_\mathrm{\lambda}=1$), we compute the scale-height $z(R)$ that fulfils
\begin{equation}
    \tau_\lambda(R) = \int_{z(R)}^{+\infty}\sum_{i} \big(\kappa^\mathrm{sca,eff}_\mathrm{\lambda,i}+\kappa^\mathrm{abs}_\mathrm{\lambda,i}\big) \rho_d^{i}(R,z)dz = 1,
\end{equation}
where we assume the dust is vertically isothermal and prescribe the dust volume density $\rho_d^{i}(R,z)$ of particle species $i$ by Equations \ref{eq:volume_density} and \ref{eq:dust_scaleheight}. We span the dust volume density $\rho_d^{i}(R,z)$ vertically from $-5H_g(R)$ to $5H_g(R)$ over 501 cells and show the emitting layer $z(R)$ in units of the gas scale-height $H_g(R)$ as a function of radius in Fig. \ref{fig:emit_layer}.

At $\lambda = 1.3~\mathrm{mm}$, thermal emission originates from 
$1$–$2H_g(R)$ in the two dust rings, but from the mid-plane elsewhere, where the disc is optically thin. Thermal emission originates from even higher layers in more turbulent discs, where dust can be stirred up more. Regardless of the turbulence, the inner dust rings have slightly higher emitting layers than the outer rings, as particles of similar sizes are more likely to be stirred up to a higher layer in the inner disc (due to their smaller Stokes numbers). A similar trend is also present at $\lambda=3~\mathrm{mm}$, though emission at longer wavelengths arises from layers closer to the mid-plane.

Since only small particles can be lifted to the emitting layer, the particle sizes $a_\mathrm{max}$ there are generally smaller than those in the mid-plane. We compare the maximum particle size $a_\mathrm{max,obs}$ above the layer where $\tau_\lambda=1$  (hereafter the maximum traceable particle size) with the maximum particle size from the mid-plane $a_\mathrm{max,real}$ in the peak of the dust ring in Table \ref{tb:amax}. The maximum particle size $a_\mathrm{max,obs}$ here is taken as the largest particle with an integrated surface density $\geq 10^{-4}~\mathrm{g~cm^{-2}}$ in the emitting layer.

In Table \ref{tb:amax} , dust larger than the maximum traceable dust grain size has been found as a result of optically thick dust rings. In such rings, only dust lifted to the emitting layers effectively contributes to emission and hence can be traced observationally, similar to the effects discussed by \citet{2020ApJ...892..136S} on spectral indices. The maximum traceable particle sizes around two dust rings are comparable, with even slightly larger dust found in the outer dust ring, regardless of dust fragility and disc turbulence. This explains the increasing \vfrag with locations of dust rings in Fig. \ref{fig:ana_sol_vfrag}. Since $a_\mathrm{max, obs}$ is more underestimated in the inner dust ring, the estimated \vfrag is biased to an even lower value there for a given $\alpha$ (Equation \ref{eq:a_max_frag}). This flat maximum traceable dust size offers a potential solution to the flat radial $a_\mathrm{max}$ profiles inferred from SED fitting \citep[e.g.][]{2021ApJS..257...14S, 2022A&A...664A.137G, 2025A&A...702A..56Z}.

\begin{table}
\centering
\caption{Table for the maximum traceable particle size $a_\mathrm{max, obs}$ at $\lambda=1.3~\mathrm{mm}$ ($a_\mathrm{max,1.3mm}$) and $3~\mathrm{mm}$ ($a_\mathrm{max,3mm}$), and the theoretical maximum particle size in the mid-plane ($a_\mathrm{max,real}$) at the peak of two dust rings (R45 and R90).}\label{tb:amax}
\begin{tabular}{ccccccccc}
\hline
\hline
 Model   & $\alpha$ & \vfrag     & \multicolumn{2}{c} {$a_\mathrm{max,1.3mm}$} & \multicolumn{2}{c} {$a_\mathrm{max,3mm}$} &\multicolumn{2}{c}{$a_\mathrm{max,real}$} \\
 & & {[$\mathrm{m/s}$]} & \multicolumn{2}{c}{{[$\mathrm{cm}$]}} & \multicolumn{2}{c}{{[$\mathrm{cm}$]}} &
 \multicolumn{2}{c}{{[$\mathrm{cm}$]}}  \\
\hline
& & & R45 & R90 & R45 & R90 & R45 & R90 \\
\hline
2 & $10^{-4}$ & 1 & 0.06 & 0.12 & 0.27 & 0.21 & 0.78 & 0.26 \\
3 & $10^{-4}$ & 2 & 0.06 & 0.11 & 0.27 & 0.47 & 3.31 & 1.18 \\
11 & $10^{-3}$ & 6 & 0.12 & 0.28 & 0.54 & 0.6 & 1.72 & 0.79 \\
12 & $10^{-3}$ & 8 & 0.12 & 0.28 & 0.53 & 1.06 &  3.34 & 1.61 \\
13 & $10^{-3}$ & 10 & 0.15 & 0.27 & 0.69 & 1.56 & 5.95 & 5.11\\
\hline
\end{tabular}
\end{table}

\subsection{Fragmentation velocity}
Fragmentation velocities in dust collisions have been widely explored in previous work. Studies using full-scale dust coagulation models, such as \dustpy, typically assume homogeneous \vfrag$\lesssim1~\mathrm{m~s^{-1}}$ or \vfrag$=10~\mathrm{m~s^{-1}}$ to study single- and multi-wavelength dust morphologies \citep[e.g.][L. Luo (in preparation)]{2021A&A...655A..18G, 2022A&A...668A.104S, 2024NatAs...8.1148U, 2024MNRAS.533.1211T, 2025ApJ...989....6K}, mostly depending on the disc turbulence. These two values are also used to study disc microphysics \citep[e.g.][]{2023A&A...674A.190D}, and the carbonaceous/non-carbonaceous dichotomy in the Solar System \citep{2023A&A...670L...5S}. A radius-dependent \vfrag has been considered to investigate the effects of snowline \citep[e.g.][]{2010A&A...513A..79B,2015A&A...584A..16P, 2017ApJ...845...68P}, across which the loss of ice-coating may reduce \vfrag \citep[e.g.][]{1997ApJ...480..647D} and affects the dust size distribution. In addition to these commonly used values at the two end, a medium value of \vfrag$=5~\mathrm{m~s^{-1}}$ is also adopted in some studies \citep[e.g.][where $\alpha$ is a few $\times 10^{-4}$]{2024A&A...688A..22L}. In population synthesis, simplified dust coagulation models, such as the two population model \citep{2012A&A...539A.148B, 2017MNRAS.469.3994B}, are used instead and to explore a wider parameter space, with \vfrag ranging from $<1$ to $>10~\mathrm{m~s^{-1}}$, and turbulence $\alpha$ ranging from $10^{-5}$ to $10^{-2}$. These studies aim to reproduce population-level trends, including the size–luminosity relation \citep{2022A&A...661A..66Z}, the continuum flux–spectral index map \citep{2024A&A...688A..81D}, and disc lifetimes/fractions \citep{2023A&A...673A..78E}. They generally prefer $\alpha\sim10^{-3}$, with higher fragmentation velocities  (\vfrag$\gtrsim 5~\mathrm{m~s^{-1}}$) required to match the observed size-luminosity relation and continuum flux-spectral index map. 

These results are generally consistent with the valid combinations of fragmentation velocities and turbulence levels that can reproduce dust rings observed by ALMA at multi-wavelengths in this work. The finding that higher turbulence ($\alpha\sim10^{-3}$) permits a slightly wider fragmentation velocity (see Table \ref{tb:model_flux}) may explain its preference in population synthesis. Though low turbulence ($\alpha\sim10^{-4}$) yields longer disc lifetime and lower stellar accretion rates \citep[see also in][]{2024MNRAS.533.1211T} that do not match current observations well, including magneto-hydrodynamic winds into a viscous model can enhance the stellar accretion rate \citep{2025ApJ...989....7T} and may mitigate this issue. We will examine whether such models can simultaneously reproduce multiple gas- and dust-phase disc properties in future studies.

\subsection{Opacity}\label{subsec:opacity}
The opacity is one of the factor that can significantly affect the physics we can derive from observations. Recent studies have shown distinct preferences in opacity models, with some preferring DSHARP opacities \citep[e.g.][]{2025MNRAS.538.2358S} while others \citep[e.g.][]{2024A&A...688A..81D} preferring opacities introduced in \citet{2010A&A...512A..15R} (hereafter Ricci10). The latter yields higher continuum millimetre fluxes due to its high absorption caused by the inclusion of amorphous carbon. To investigate whether variations in the assumed opacities can alter the detectability of weakly emitted dust rings and hence the valid combinations of \vfrag and $\alpha$, we replace the DSHARP opacities in Section \ref{subsec:RT} with Ricci10 opacities, and summarise the resulting disc properties in Table. \ref{tb:model_flux_ricci10}.

Adopting the Ricci10 opacities does indeed increase the continuum fluxes and, in some models, improves the detectability of dust rings. For example, the outer dust ring at Band 3 becomes tentatively detected (instead of being undetected) in Model 1, and becomes robustly detected (instead of tentatively detected) in Model 10. However, except for Model 10, which becomes a new valid model, models that are invalid with DSHARP opacities remain invalid with Ricci10 opacities. This demonstrates that the valid combinations of $\alpha$ and \vfrag shown in Section \ref{subsubsec:valid_models} are not sensitive to the adopted opacities. 

However, our numerical simulations assume a dust bulk density of $1.67~\mathrm{g~cm^{-3}}$, which is higher than the bulk density of porous grains in the Ricci10 opacity. This treatment over-estimates the Stokes number of particles at a given size (Equation \ref{eq:stokes}), leading to more dust being trapped around the dust rings. A more consistent treatment on the dust bulk density would be helpful to confirm this conclusion.

\begin{table*}
\caption{Same as Table \ref{tb:params} but adopting Ricci10 opacities \citep{2010A&A...512A..15R}.} \label{tb:model_flux_ricci10}
\begin{tabular}{cccccccccccccc}
\hline
\hline
(1) & (2) & (3) & (4) & (5) & (6) & (7) & (8) & (9) & (10) & (11) & (12) & (13) & (14) \\
Model &  $\alpha$& $v_\mathrm{frag}$ & $R_\mathrm{1.3mm}^\mathrm{1myr}$ & $R_\mathrm{3mm}^\mathrm{1myr}$ &  
$R_\mathrm{1.3mm}^\mathrm{3myr}$ & $R_\mathrm{3mm}^\mathrm{3myr}$ & 
$F_\mathrm{1.3mm}^\mathrm{1myr}$ &$F_\mathrm{3mm}^\mathrm{1myr}$ &$F_\mathrm{1.3mm}^\mathrm{3myr}$ & $F_\mathrm{3.0mm}^\mathrm{3myr}$ & $\alpha_\mathrm{1.3-3mm}^\mathrm{1myr}$ & $\alpha_\mathrm{1.3-3mm}^\mathrm{3myr}$ & valid model? \\
 &  & [$\mathrm{m~s^{-1}}$] & in/out & in/out  & in/out & in/out & [mJy] & [mJy] & [mJy] & [mJy] & & & \\
\hline
1 & $10^{-4}$ & $0.5$ &\cmark\cmark &\cmark\bqm & \cmark\cmark  & \cmark\cmark  & 332.17  &28.38 & 267.41 & 20.89 & 2.94 & 3.04 & \\
2 & $10^{-4}$ & $1$ & \cmark\cmark & \cmark\cmark & \cmark\cmark & \cmark\cmark & 283.29 & 23.65 & 180.70 & 23.06 & 2.97 & 2.46 & \cmark \\
3 & $10^{-4}$ & $2$ & \cmark\cmark & \cmark\cmark & \cmark\cmark & \cmark\cmark & 226.18 & 21.69 & 138.15 & 18.72 & 2.80 & 2.39 & \cmark \\
4 & $10^{-4}$ & $4$ & \cmark\cmark & \cmark\cmark & \cmark\cmark & \cmark\bqm & 155.72 & 15.19 & 74.09 & 9.18 & 2.78 & 2.50 & \\
5 & $10^{-4}$ & $6$ & \cmark\cmark & \xmark\xmark & \cmark\cmark & \xmark\xmark & 77.43 & 5.40 & 16.16 & 1.68 & -- & -- &  \\
6 & $10^{-4}$ & $8$ & \cmark\cmark & \xmark\xmark & \xmark\bqm & \xmark\xmark & 74.16 & 5.31 & 10.57 & 1.10 & -- & -- & \\
7 & $10^{-4}$ & $10$& \cmark\cmark & \xmark\xmark & \xmark\bqm & \xmark\xmark & 74.16 & 5.17 & 10.57 & 1.08 & -- & -- & \\
8 & $10^{-3}$ & $1$& \cmark\bqm & \xmark\xmark & \cmark\bqm & \xmark\xmark & 291.15 & 18.90 & 264.34 & 14.33 & -- & -- & \\
9 & $10^{-3}$ & $2$& \cmark\cmark & \cmark\bqm & \cmark\cmark & \cmark\bqm & 541.93 & 35.51 & 465.69 & 31.84 & 3.26 & 3.21 & \\
10 & $10^{-3}$ & $4$& \cmark\cmark & \cmark\cmark & \cmark\cmark & \cmark\cmark & 399.71 & 40.87 & 298.84 & 32.02 & 2.73 & 2.67 & \textcolor{red}{\cmark}\\
11 & $10^{-3}$ & $6$& \cmark\cmark & \cmark\cmark & \cmark\cmark & \cmark\cmark & 324.60 & 36.99 & 259.83 & 31.44 & 2.60 & 2.53 & \cmark \\
12 & $10^{-3}$ & $8$& \cmark\cmark & \cmark\cmark & \cmark\cmark & \cmark\cmark & 305.17 & 34.48 & 242.64 & 28.85 & 2.61 & 2.55 & \cmark\\
13 & $10^{-3}$ & $10$& \cmark\cmark & \cmark\cmark & \cmark\cmark & \cmark\cmark &  294.85 & 32.76 & 232.71 & 28.33 & 2.63 & 2.52 & \cmark \\
14 & $10^{-3}$ & $12$& \cmark\cmark & \cmark\cmark & \cmark\cmark & \cmark\bqm &  240.53 & 24.14 & 131.34 & 15.17 & 2.75 & 2.58 & \\
\hline

\end{tabular}
\end{table*}

\subsection{Bouncing barriers}\label{subsec:discussion_dust_prop} 
\begin{figure*}
    \centering
    \includegraphics[width=0.99\linewidth]{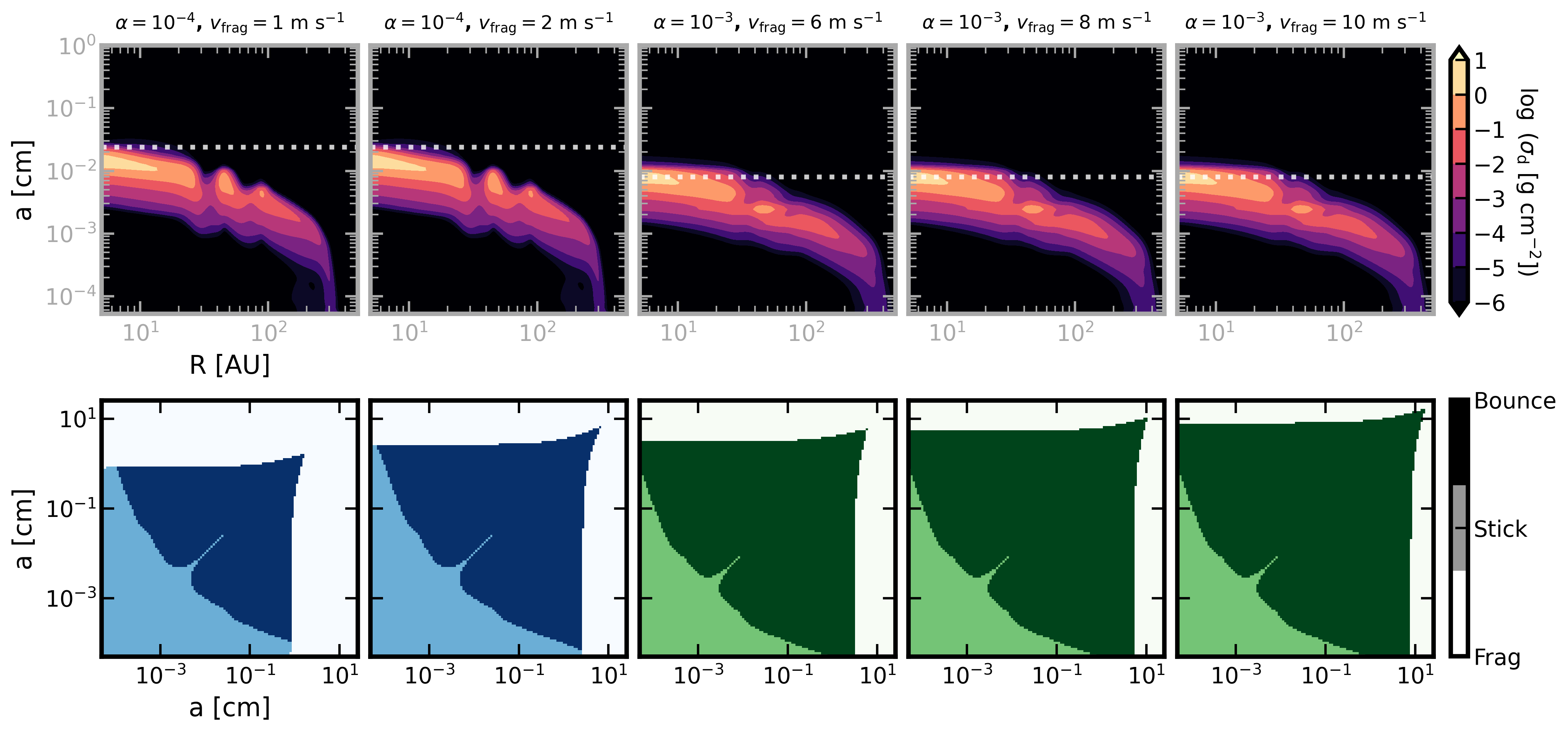}
    \caption{Results from numerical models for Models 2-3 and Models 11-13 with bouncing considered. Upper panels: dust size distributions for bouncing-integrated Models 2-3 and 11-13 (from left to right). The white dotted line in each panel indicates the maximum particle size that bouncing-limited grains can grow in the innermost disc. Lower panels: outcome of collisions between particles in the innermost disc ($R=$5 au), for the same models and in the same order as the upper panels.}
    \label{fig:bouncing}
\end{figure*}

Previous studies have shown that bouncing is one of the key barriers to dust growth from $\mu \mathrm{m}$ to mm/cm size \citep[e.g.][]{2010A&A...513A..57Z}. It occurs when the relative colliding velocity exceeds the bouncing velocity $v_b$ but is still below the fragmentation velocity \vfrag. Particles colliding within this range bounce off instead of sticking or fragmenting, preventing them from being destroyed through catastrophic fragmentation. This results in a quasi-monodisperse size distribution \citep[e.g.][]{2010A&A...513A..57Z, 2024A&A...682A.144D} when no centimetre dust is introduced initially \citep{2012A&A...540A..73W}. Such a distribution is proposed as an efficient mechanism to retain dust and keep discs optically thick, to explain the universal size-luminosity relation \citep[e.g.][]{2025arXiv250706298Q}. 

In this section, we prescribe a bouncing velocity in our models, following \citet{2010A&A...513A..57Z},
and investigate whether bouncing has impacts on the five valid models in reproducing the general features of multi-wavelength observations. The bouncing velocity is defined as 
\begin{equation}\label{eq:bounce}
    v_\mathrm{b}= \sqrt{\frac{5\pi a_0 F_\mathrm{roll}}{m_\mu}}.
\end{equation}
Here $a_0$ is the radius of dust monomer, which is assumed to be $0.5~\mathrm{\mu m}$ in our models (see Table \ref{tb:params}). $F_\mathrm{roll}$ is the rolling force\footnote{Rolling force is the force required for one dust monomer to roll over another \citep[see][]{1997ApJ...480..647D}} and $m_\mu=(m_1m_2)/(m_1+m_2)$ is the reduced mass of two colliding particles with masses $m_1$ and $m_2$, respectively. We adopt $F_\mathrm{roll}=10^{-4}$ dyne, a value determined by experiments \citep{1999PhRvL..83.3328H}.

The results for the five valid models incorporating bouncing are shown in Fig. \ref{fig:bouncing}. As expected, all the five models show narrow dust size distributions and none of them form particles larger than 1 mm. These smaller grains are better coupled with gas and better retained in the disc instead of drifting inwards rapidly. This enhances the dust surface density, particularly in the inner region, compared to the case where bouncing is not introduced. This dust size distribution, shown in the upper panel of Fig. \ref{fig:bouncing}, is not sensitive to the fragmentation velocity but weakly depends on the disc turbulence levels. The bouncing-limited dust can grow to a slightly larger size in a low-turbulence disc, where turbulence-dominated relative velocities are less likely to exceed the bouncing threshold. This is well aligned with the analytical solution of the bouncing-limited dust size $a_b$ presented in \citet{2024A&A...682A.144D}.
\begin{equation}
    a_{b}= \bigg( \frac{5}{\pi} \frac{\Sigma_g a_0 F_\mathrm{roll}}{\alpha c_s^2 \rho_s}\bigg)^{1/4},
\end{equation}
where $a_b \propto \alpha^{-0.25}$. 

We show the outcome of pair-wise collisions in the innermost disc in the lower panel of Fig. \ref{fig:bouncing} after comparing dust relative colliding velocities with bouncing and fragmentation velocities for five valid models. It is evident that without the presence of large particles initially, the maximum colliding pairs allowed for sticking is the equal-mass collision between particles with sizes of $10^{-3}$--$10^{-2}$ cm. We highlight the values as dotted lines in the upper panel and this is consistent with the maximum dust sizes in numerical models.

When larger particles ($\gtrsim 10^{-2}$ cm) are originally presented in the disc, they can continue growing to millimetre and even centimetre sizes by sticking to smaller particles, after which they enter the bouncing regime again. The size they can grow to before triggering bouncing is a function of turbulence and the fragmentation velocity. A lower turbulence and higher fragmentation velocity, like Model 3 ($\alpha=10^{-4}$, \vfrag=$2~\mathrm{m~s^{-1}}$), is preferred for the optimal dust growth.

\begin{figure}
    \centering
    \includegraphics[width=0.95\linewidth]{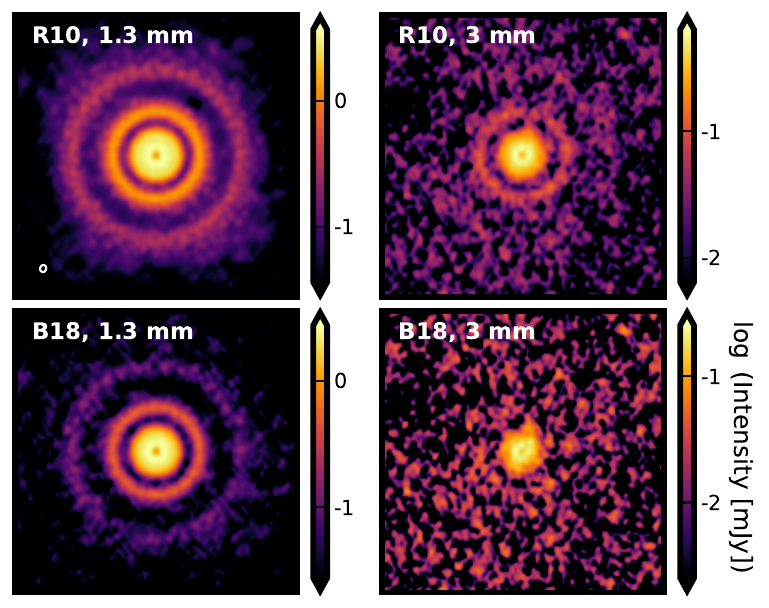}
    \caption{Synthetic observations of Model 3+bouncing adopted Ricci10 (upper, abbreviated as ``R10'') and DSHARP (lower, abbreviated as ``B18'') opacities at Band 6 ($\lambda=1.3$ mm, left) and Band 3 ($\lambda=3$ mm, right). Images are shown in log-scale for better illustration of dust features. The small inner cavity is due to the inner boundary being set at 5 au. A representative beam size is plotted at the lower left corner on the upper left panel.}
    \label{fig:bounce_obs}
\end{figure}

However, without introducing large particles, all models where dust growth is limited by bouncing have difficulties in reproducing the general features in multi-wavelength observations, due to the lack of large particles, irrespective of the adopted opacities. We show synthetic observations generated from Model 3 ($\alpha=10^{-4}$, \vfrag=$2~\mathrm{m~s^{-1}}$)+bouncing in Fig. \ref{fig:bounce_obs}. Mock observations can reproduce two dust rings in Band 6 (regardless of the choice of opacity). However, at longer wavelengths, the presence of larger particles is required to enable the detectability of outer dust rings. The weak emission due to insufficient large dust cannot be compensated for by the high optical depth caused by bouncing. 

We speculate that porous dust grains may provide a solution to the detectability of outer dust rings at longer wavelengths in this bouncing-limited regime. Porosity can boost the absolute dust size, which is a quantity directly probed by observations, at a fixed mass; while porosity lowers down the density of aggregates and hence decreases the Stokes number (Equation \ref{eq:stokes}), reducing the relative collision velocity and promoting further dust growth. However, porous grains can have lower opacity than compact grains at millimetre, depending on the filling factor. The net effects on the dust ring brightness by porous dust require further studies to confirm.

\subsection{Properties of dust rings}\label{subsec:discussion_dust_rings}

\begin{figure*}
    \centering
    \includegraphics[width=0.954\linewidth]{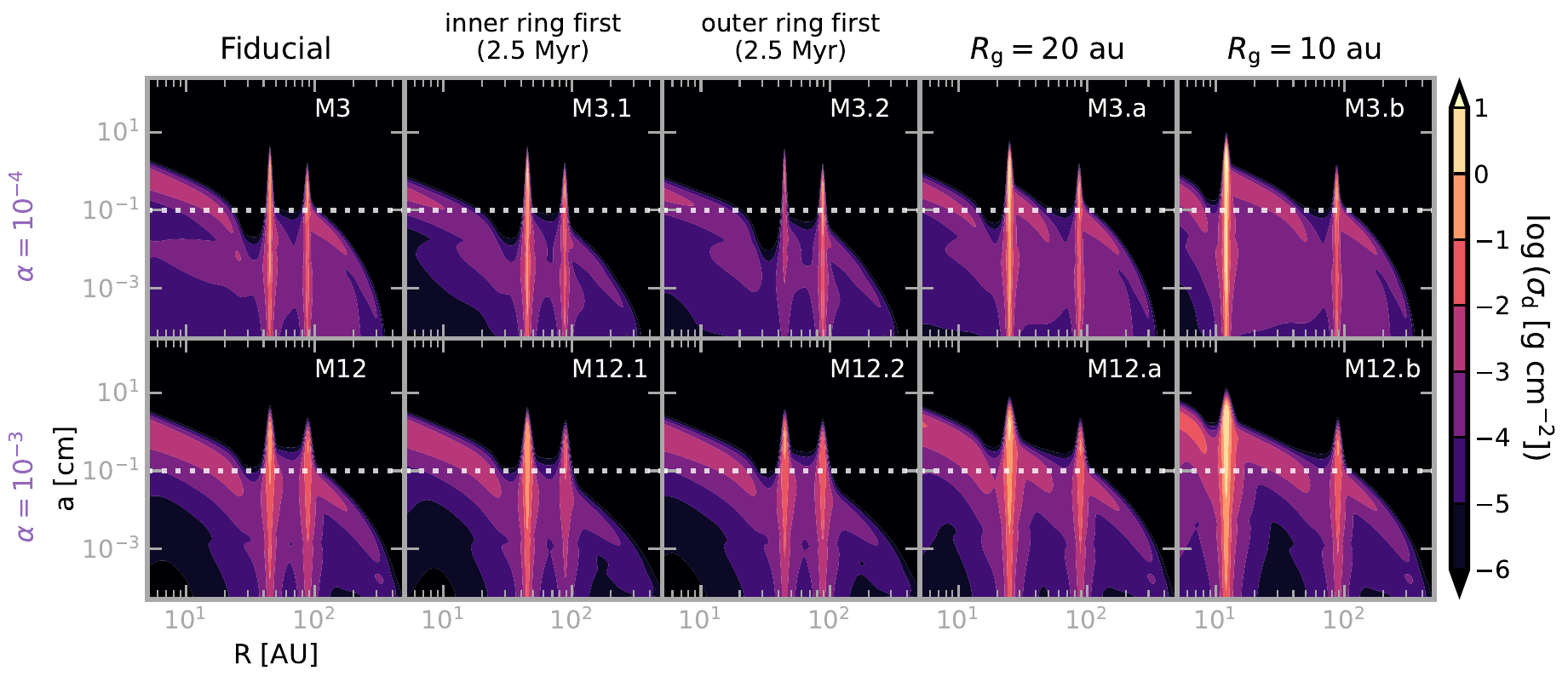}
    \caption{The dust size distribution along radius for models shown in Table \ref{tb:order&separation}. The first column shows the fiducial models Models 3 ($\alpha=10^{-4}$, \vfrag$=2~\mathrm{m~s^{-1}}$) and 12 ($\alpha=10^{-3}$, \vfrag$=8~\mathrm{m~s^{-1}}$). The second and third columns show the modified models of Model 3/12 with inner and outer dust rings appearing first, respectively. The fourth and fifth columns show the modified models with dust rings at different separations. The white dotted line in each panel indicates the grain size $a=0.1~\mathrm{cm}$. Unless otherwise specified, models are shown at $t=1~\mathrm{Myr}$.}
    \label{fig:dust_ring_numerical}
\end{figure*}

In Section \ref{sec:numerical}, we introduce the pressure bumps at the same time and only consider one value of dust ring separation. In practice, dust pressure bumps do not necessarily appear simultaneously, and they can have different separations. These factors regulate dust supply from further out in the disc, and the efficiency of local dust accumulation, and thereby alter the intensity ratios of dust rings. In this section, we run additional 8 models to briefly investigate how the order of appearance of pressure bumps, and their separations, affect dust retention and relevant observables. Our new models are based on the valid Models 3 ($\alpha=10^{-4}$, \vfrag=$2~\mathrm{m~s^{-1}}$) and 12 ($\alpha=10^{-3}$, \vfrag=$8~\mathrm{m~s^{-1}}$) in Table \ref{tb:model_flux}. We assume the DSHARP opacities when generating the radiative transfer model. We limit our discussion to the simplest case, where only two dust pressure bumps are present in the disc. All the models discussed in this section are summarised in Table \ref{tb:order&separation}. Their size distributions at $t=1$/$2.5$ Myr are shown in Fig. \ref{fig:dust_ring_numerical}.

\begin{table*}
\caption{Disc properties of modified models 3 and 12 with various dust ring setup. Column 2 is for the turbulence level and Column 3 is for the fragmentation velocity. Column 4 is the locations of pressure bumps $R_g$ for the inner and outer dust rings. Column 5 is the time when the pressure gradient becomes positive ($\mathrm{d}\ln P/\mathrm{d}\ln R>0$) for the inner and outer dust rings. Column 6 is the theoretical total dust disc mass. Columns 7 and 8 are the continuum fluxes at $\lambda=1.3$ and $\lambda=3$ mm. Column 9 is the spectral index computed from columns 7 and 8. All the quantities reported in columns 6--9 are measured at 2.5 Myr for the appearance orders section and at 1 Myr for the separations section.}\label{tb:order&separation}
\begin{tabular}{ccccccccccc}
\hline
\hline
(1) & (2) & (3) &  \multicolumn{2}{c}{(4)} & \multicolumn{2}{c}{(5)} & (6) & (7) & (8) & (9)\\
Model & $\alpha$ & \vfrag & \multicolumn{2}{c}{$R_\mathrm{g}$} & \multicolumn{2}{c}{$t_\mathrm{appear}$} & $M_{d,\mathrm{tot}}$ & $F_\mathrm{1.3mm}$ & $F_\mathrm{3mm}$ & $\alpha_\mathrm{1.3-3mm}$ \\
 &  & [$\mathrm{m~s^{-1}}$] & \multicolumn{2}{c}{[au]} & \multicolumn{2}{c}{[Myr]} & [$M_\oplus$] & [mJy] & [mJy] & \\
\hline
 &  &  & inner & outer & inner & outer & & & & \\
 \hline
\multicolumn{11}{c}{Appearance orders}\\
\hline

 3.1 & $10^{-4}$ & 2 & 35 & 70 & 0 & 1  & 257.56 & 69.76 & 8.70 & 2.48\\
 3.2 & $10^{-4}$ & 2 & 35 & 70 & 1.1 & 0 & 171.73 & 53.80 & 7.04 & 2.43\\

 12.1 & $10^{-3}$ & 8 & 35 & 70 & 0 & 0.9 & 244.83 & 85.89 & 8.97 & 2.70\\
 12.2 & $10^{-3}$ & 8 & 35 & 70 & 1.1  & 0 & 161.35 & 119.09 & 11.73  & 2.77 \\
 \hline
 \multicolumn{11}{c}{Separations}\\
 \hline
3   & $10^{-4}$ & 2 & 35 & 70 & 0 & 0 & 255.91 & 99.66  & 9.73 & 2.78 \\
3.a & $10^{-4}$ & 2 & 20 & 70 & 0 & 0 & 303.19 & 96.79 & 9.38 & 2.79 \\
3.b & $10^{-4}$ & 2 & 10 & 70 & 0 & 0 & 337.82 & 90.96 & 9.01 & 2.76 \\
12  & $10^{-3}$ & 8 & 35 & 70 & 0 & 0 & 241.33 & 149.07 & 13.48 & 2.87\\
12.a & $10^{-3}$ & 8 & 20 & 70 & 0 & 0 & 284.31 & 134.64 & 14.30 & 2.68 \\
12.b & $10^{-3}$ & 8 & 10 & 70 & 0 & 0 & 304.69 & 112.83 & 13.35 & 2.55 \\
\hline
\end{tabular}
\end{table*}

\subsubsection{Order of appearance for dust rings}\label{subsec:discussion_dust_order}
In this section, we introduce one pressure bump at the start of the simulations, and the other later. We define the time when the second pressure bump has a positive pressure gradient ($\mathrm{d}\ln P/\mathrm{d}\ln R>0$) as the appearance time $t_\mathrm{appear}$ in Table \ref{tb:order&separation}. As one of the pressure bumps is introduced at $\sim 1$ Myr, we generate synthetic observations later than $2$ Myr to allow dust trapped in the second pressure bump to grow.

Discs with the inner pressure bump introduced first retain $50$ per cent more dust mass at $t=2.5~\mathrm{Myr}$. This is because pressure bumps can only possibly trap particles drifting in from larger radii, and an inner pressure bump has larger dust supply than an outer one. If the outer pressure bump appears first and the inner one has not, large grains formed in the outer pressure bump can still be lost from the disc when they fragment to smaller particles and fail to be trapped locally. However, the enhanced dust reservoirs in discs where the inner pressure bump appears first do not translate proportionally into brighter millimetre continuum emission, since much of additional mass is ``hidden'' in the optically thick dust rings.

\begin{figure}
    \centering
    \includegraphics[width=1.00\linewidth]{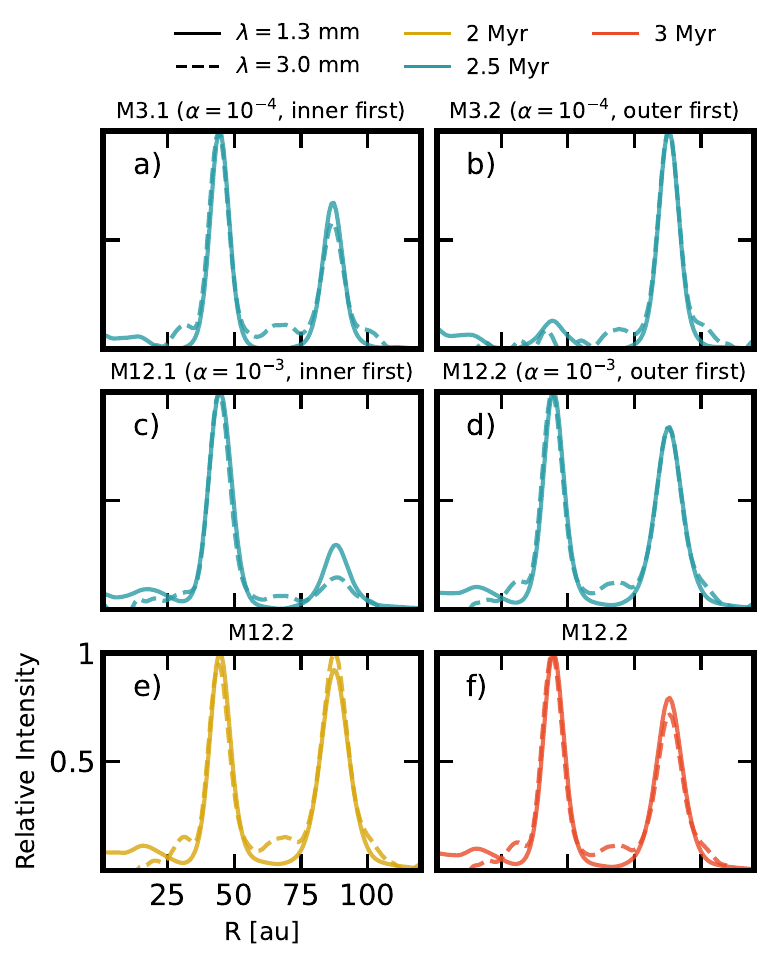}
    \caption{Relative intensity of dust rings introduced at different time at $\lambda=1.3$ mm (solid lines) and $3$ mm (dashed lines). The relative intensity is read from synthetic observations based on models summarised in Table \ref{tb:order&separation} and Fig. \ref{fig:dust_ring_numerical}. Panels a-d are generated from numerical models at $2.5$ Myr (blue). Panels e and f are generated for models at $2$ (yellow) and $3$ Myr (orange), respectively. }
    \label{fig:appearance}
\end{figure}

The appearance order also affects disc morphologies. Inner dust rings introduced later appear fainter than the first appeared outer dust ring, and the case is more extreme here as they are introduced so late that they even remain undetected ($<3\sigma$) at Band 3 ($\lambda=3$ mm, see Fig. \ref{fig:appearance}) in Model 3.2. Such trend is not common in current millimetre observations of protoplanetary discs, most of which have decreasing dust ring brightness with radius \citep[with very limited exceptions, such as the circumbinary disc V4046 Sgr;][]{2022MNRAS.510.1248M, 2025ApJ...984L...9C}. Though an inner dust ring introduced later could become brighter than the outer one, such as Model 12.2 at $\gtrsim 2$ Myr (Fig. \ref{fig:appearance}), we do not consider this as a valid model (as discussed in Section \ref{subsec:result_intensity_ratio}), because the intensity ratio of the inner to outer rings changes substantially over a relatively short period $\lesssim 1$ Myr--from a brighter outer dust ring to a brighter inner dust ring. (panels d-f in Fig. \ref{fig:appearance}). These models suggest that inner dust rings have to appear first, or at least not long ($\lesssim 1$ Myr) after the appearance of outer rings, in order for the inner ring to appear brighter (as found in the majority of ALMA observations). 

\subsubsection{Separation of dust rings}

To investigate the impact of separations of pressure bumps on dust retention, we fix the location of the outer pressure bump at $70$\,au, and move the inner pressure bump from $\sim 45$\,au to $\sim 25$\,au and $\sim 10$\,au for models 3.a, 12.a and 3.b, 12.b, respectively. Both dust rings are introduced at the start of simulations, and we keep the pressure gradients for both pressure bumps similar for all the models explored here.

\begin{figure}
    \centering
    \includegraphics[width=0.9\linewidth]{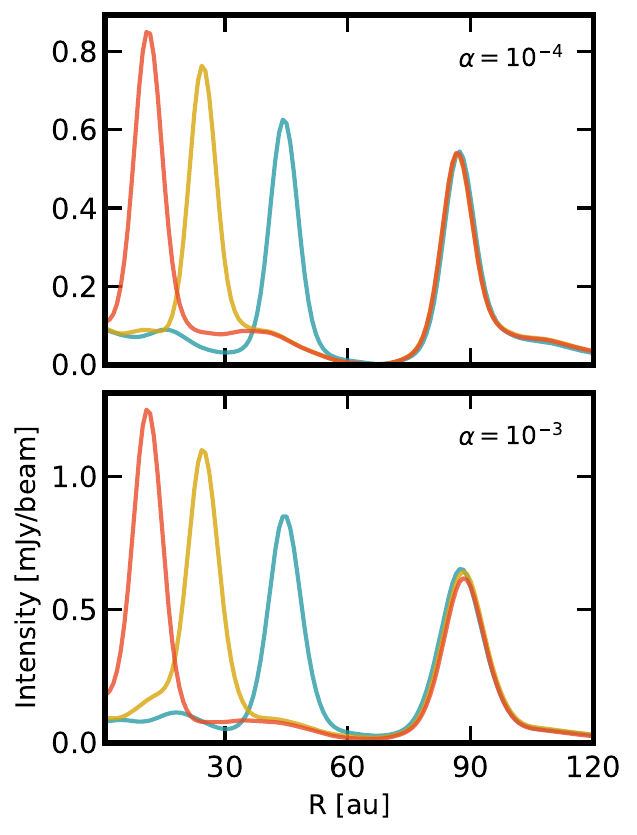}
    \caption{Intensity of models with pressure bumps separated by 45 (blue), 65 (yellow) and 80 (red) au at $\lambda=1.3~\mathrm{mm}$ and $t=1~\mathrm{Myr}$ read from the synthetic observations generated from Model 3, 3.a and 3.b for discs with $\alpha=10^{-4}$ (upper panel), and from Model 12, 12.a and 12.b for discs with $\alpha=10^{-3}$ (lower panel) in Fig. \ref{fig:dust_ring_numerical}.}
    \label{fig:sep_intensity}
\end{figure}

Models with more widely separated dust rings retain dust better, particularly for the inner pressure bumps, and are more successful in maintaining higher intensity ratios as shown in Fig. \ref{fig:sep_intensity}, regardless of the turbulence level. This enhanced dust retention in the inner disc is due to the shorter time-scale for dust growth at smaller radii. The dust growth time-scale is derived as 
\begin{equation}\label{eq:growth_timescale}
    \frac{\mathrm{d}a}{\mathrm{d}t}=\frac{\rho_d}{\rho_s}\Delta v_\mathrm{rel}
\end{equation}
in \citet{2001A&A...378..180K}. Here, $a$ is the dust size, $\rho_d$ is the dust volume density and $\rho_s$ is the dust bulk density. Assuming that turbulent motion dominates the dust velocity and that $\epsilon=\Sigma_d/\Sigma_g$ \citep{2012A&A...539A.148B},  Equation \ref{eq:growth_timescale} gives
\begin{equation}
    \frac{\mathrm{d}a}{\mathrm{d}t}= \sqrt{\frac{3\pi}{8}}a \epsilon \Omega_k,
\end{equation}
indicating that dust grows more rapidly as it drifts inwards. The short growth time-scale makes it easier for dust to grow to the critical size that can be trapped around the pressure bumps.

When dust sizes are limited by radial drift, which is true for dust between two rings explored here, competition exists between radial drift and dust growth. This indicates that dust which fails to be trapped by the outer pressure bump may not be able to grow to sufficiently large sizes that can be trapped by the inner pressure bump, if the growth timescale is longer than the drift timescale between two close pressure bumps. This may lead to a faint inner dust ring when two rings are close together, or even a faint middle ring sandwiched between two rings, as seen in the AA Tau system \citep[B72 in ][]{2025ApJ...984L...9C}.

%% file: mainbody/conclusion.tex
In this paper, we investigate how the joint choice of disc turbulence levels $\alpha$ and dust fragility \vfrag can best reproduce recent multi-wavelength ALMA observations of protoplanetary discs, using both analytical and numerical models. In the analytical model, we infer combinations of $\alpha$ and \vfrag required to form the maximum dust sizes derived for three out of five MAPS targets, adopting stellar and disc parameters from studies based on MAPS observations. In the numerical models, we use \dustpy to explore combinations of $\alpha$ and \vfrag that can reproduce the general trends of multi-wavelength observations: a. consistent detections of dust rings at $\lambda=1.3$ (Band 6) and $3$ mm (Band 3); b. dust ring intensity ratios that do not change substantially over Myr timescales and across wavelengths; and c. the spectral indices between Bands 3 and 6 within the reasonable change from previous studies ($2\lesssim \alpha_\mathrm{1.3-3mm}\lesssim 4$). We summarise our results as follows:
\begin{itemize}
    \item Analytical and numerical results consistently suggest that dust fragility is positively correlated with disc turbulence levels. Resilient dust (\vfrag =$6$--$10~\mathrm{m~s^{-1}}$) is required in more turbulent discs ($\alpha=10^{-3}$) and fragile dust (\vfrag =$1$--$2~\mathrm{m~s^{-1}}$) is required in less turbulent discs ($\alpha=10^{-4}$), indicating discs must have low turbulence ($\alpha\sim 10^{-4}$) if dust is fragile. However,  very fragile dust (\vfrag$<1~\mathrm{m~s^{-1}}$) struggles to reproduce multi-wavelength observations. These results are robust to the choice of opacities (DSHARP and Ricci10).
    \item Reproducing multi-wavelength observations requires dust rings to be optically thick at $\lambda=1.3$ and $3~\mathrm{mm}$ over Myr time-scales. Standard SED fitting that neglects the dust vertical distribution tends to underestimate the maximum dust size, especially at small radii, resulting in a radially flat maximum grain size profile.
    \item A mono-disperse, bouncing-limited dust size distribution struggles to reproduce multi-wavelength observations, due to the lack of sufficient millimetre dust in the outer discs. 
    \item The order of appearance and the separations of pressure bumps have impacts on dust retention and millimetre disc morphologies. Dust rings which decrease in brightness with increasing radius (as is commonly observed) favour models in which the inner dust ring forms earlier. 
    \item Our structured disc models predict that the spectral index between $\lambda=1.3$ and $3~\mathrm{mm}$ decreases over time as a result of changes in the dust size distribution.
    \end{itemize}

%% file: appendix/appendix1.tex
\section{Overview of invalid models}\label{sec:appendix_invalid}
\begin{figure*}
    \centering
    \includegraphics[width=0.73\linewidth]{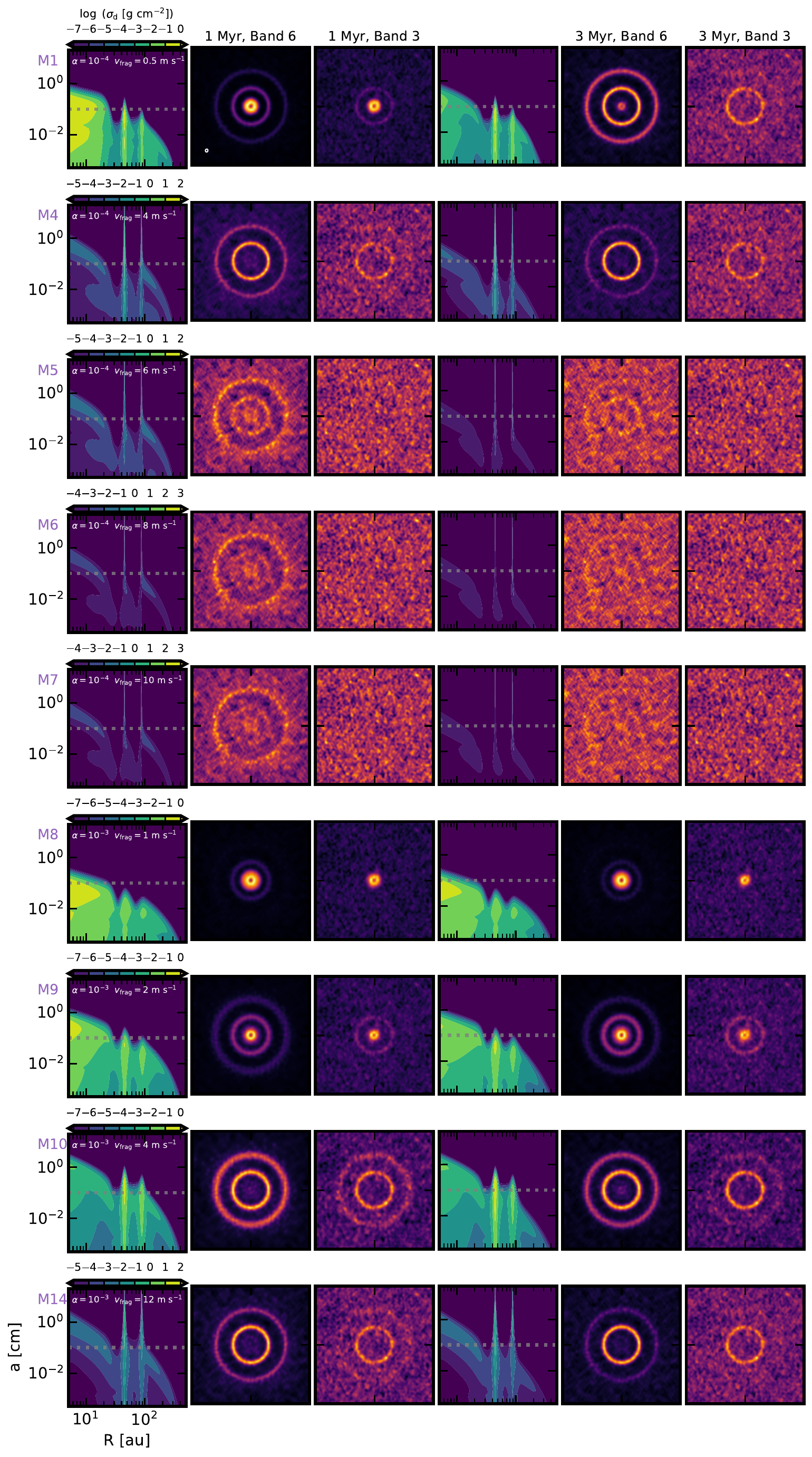}
    \caption{Same as Fig. \ref{fig:valid_model} but for invalid models. The small inner cavity in some synthetic observations is due to the inner boundary being set at 5 au.}
    \label{fig:invalid_model}
    \label{lastpage}
\end{figure*}

\begin{figure}
    \centering
    \includegraphics[width=0.9\linewidth]{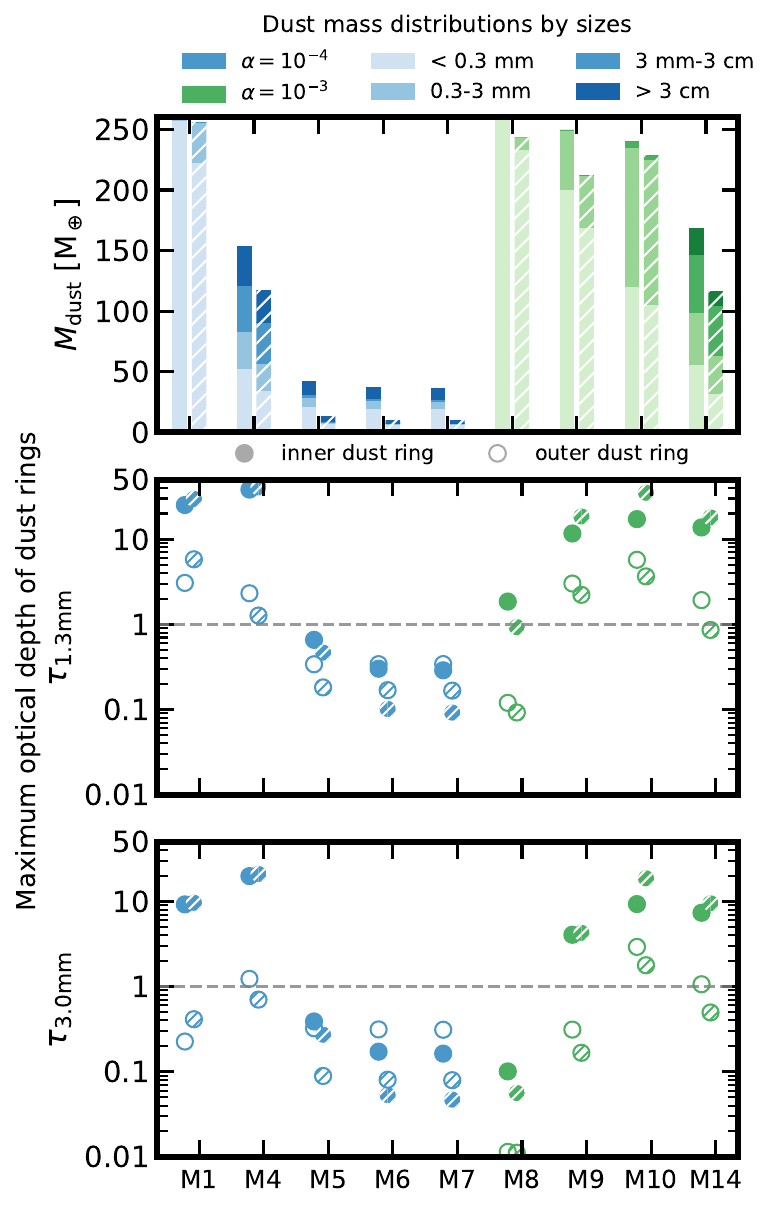}
    \caption{Same as Fig. \ref{fig:optical_depth} but for invalid models. The non-hatched and hatched bars/dots are for measurements at 1 and 3 Myr, respectively. The dashed grey lines in the middle and lower panels indicates $\tau_\nu=1$.}
    \label{fig:invalid_optical_depth}
    \label{lastpage}
\end{figure}

The overview of all the invalid models are shown in Fig. \ref{fig:invalid_model}. The majority of them become invalid due to the failure in reproducing the outer dust rings in Band 3 at 3 Myr or at both 1 and 3 Myr. The fundamental cause is the insufficiency of dust with proper sizes, leading to the low opacity. This can be found in the upper panel of Fig. \ref{fig:invalid_optical_depth}, where we show the size composition of dust for each model. In the middle and lower panels of the same figure, we show the optical depth around the dust rings of each model. The low optical depth ($<1$) of the outer dust rings sustains during 1--3 Myr.

